\documentstyle[12pt,epsf]{article}

   \def\unlock{\catcode`@=11}

   \unlock

   \def\gsim{\mathrel{\mathpalette\@versim>}}

   \def\@versim#1#2{\vcenter{\offinterlineskip
        \ialign{$\m@th#1\hfil##\hfil$\crcr#2\crcr\sim\crcr } }}

\newcommand{\zp}[3]{Z. Phys.\ {\bf C#1}, #2 (19#3)}
\newcommand{\pl}[3]{Phys.\ Lett.\ {\bf #1B}, #2 (19#3)}
\newcommand{\np}[3]{Nucl.\ Phys.\ {\bf B#1}, #2 (19#3)}
\newcommand{\prd}[3]{Phys.\ Rev.\ {\bf D#1}, #2 (19#3)}
\newcommand{\prl}[3]{Phys.\ Rev.\ Lett.\ {\bf #1}, #2 (19#3)}
\newcommand{\sjnp}[3]{Sov.\ J.\ Nucl.\ Phys.\ {\bf #1}, #2 (19#3)}
\newcommand{\jetp}[3]{Sov.\ Phys.\ JETP\ {\bf #1}, #2 (19#3)}
\newcommand{\prep}[3]{Phys.\ Rep.\ {\bf C#1}, #2 (19#3)}
\newcommand{\yad}[3]{Yad.\ Fiz.\ {\bf #1}, #2 (19#3)}
\newcommand{\zh}[3]{Zh.\ Eksp.\ Teor.\ Fiz. {\bf #1}, #2 (19#3)}

\newcommand{\beq}{\begin{equation}}
\newcommand{\eeq}{\end{equation}}
\newcommand{\bea}{\begin{eqnarray}}
\newcommand{\eea}{\end{eqnarray}}

\newcommand{\pcol}{p_{_{||}}}
\newcommand{\ptp}{p_\perp}
\newcommand{\kt}{k_\perp}
\newcommand{\kta}{k_{a\perp}}
\newcommand{\ktb}{k_{b\perp}}
\newcommand{\kti}{k_{i\perp}}
\newcommand{\qta}{q_{a\perp}}
\newcommand{\qtb}{q_{b\perp}}
\newcommand{\ptm}{p_{\perp\rm min}}

\newcommand{\ra}{\rightarrow}
\newcommand{\lra}{\leftrightarrow}
\renewcommand{\a}{\alpha}
\newcommand{\s}{\sigma}
\newcommand{\hsig}{\hat \sigma}
\newcommand{\hs}{\hat s}
\newcommand{\th}{\hat t}
\newcommand{\nn}{\nonumber}
\newcommand{\un}{\underline}

\begin{document}

\begin{titlepage}

\hspace*{\fill}\parbox[t]{2.8cm}{DESY 95-023\\DFTT 13/95\\ February 1995}

\vspace*{1cm}

\begin{center}
\large\bf
An introduction to the perturbative QCD pomeron and to jet physics at large
rapidities
\end{center}

\vspace*{0.5cm}

\begin{center}
Vittorio Del Duca \\
Deutsches Elektronen-Synchrotron \\
DESY, D-22603 Hamburg , GERMANY
\end{center}

\vspace*{0.5cm}

\begin{center}
\bf Abstract
\end{center}

\noindent
In these lectures we discuss
the Balitsky, Fadin, Kuraev, and Lipatov (BFKL) theory, which resums the
leading logarithmic contributions to the radiative corrections to parton
scattering in the high-energy limit,
and we apply it to hadronic two-jet production at large rapidity intervals.
\end{titlepage}

\baselineskip=0.8cm

\section{Introduction}

This is an expanded version of a few lectures I gave at the
Universita` di Salerno and Torino, Italy, on the perturbative QCD pomeron
and on jet physics at large rapidity intervals. The outline of the lectures
is as follows: in the Introduction the parton-model and the factorization
pictures, propedeutic to any calculation of strong-interaction processes,
are sketched; in sect.~2 two-jet production at
hadron colliders at the leading order in $\a_s$ is discussed, by
examining the parton kinematics and dynamics in the exact configurations and
in the large-rapidity limit; in sect.~3 higher-order corrections
to two-parton scattering are considered in the limit of a strong rapidity
ordering of the final-state partons. In resumming the leading logarithmic
contributions to the radiative corrections, the BFKL equation describing
the gluon-ladder evolution in transverse momentum is introduced; in sect.~4
the BFKL formalism is applied to the description of inclusive two-jet
production at large rapidity intervals.

\subsection{Jet Production and Factorization}
\label{sec:oneone}

The importance of jets in hadronic collisions has been realized since the
conception of the parton model, where jets arise from the scattering
between the constituent partons of the colliding hadrons. On the experimental
side, hadronic jets
were then actively searched at the CERN ISR collider, but their existence
was demonstrated only at the CERN SPS collider \cite{ua}, by
applying a cut in the total scalar transverse energy of the event. This
strongly reduced the soft-hadron background of low transverse-energy partons
due to the ``underlying event", i.e. to the soft scattering between the
spectator partons, and left mainly high transverse-energy jets
\footnote{An introduction to the experimental discovery and to the theory of
hadronic jets is given in ref.\cite{ell}.}

In the parton model the scattering
process is factorized into two regions: $i$) a short-distance region,
characterized by a hard scale $Q$, of the order of the jet transverse energy
$\sim 10$ GeV and thus typical distances of
about $10^{-2}$ fm, which describes the primary scattering between the
partons;
$ii$) a long-distance region, characterized by a hadronization scale
$\sim \Lambda_{QCD}$ and distances of about 1 fm, which describes how
the scattering partons split from the parent hadrons and how
the scattered partons hadronize.

To see intuitively how this factorization comes about in the
parton-model picture, let us imagine to sit in the center-of-mass frame of
the hadron scattering. Then we observe two kinematic effects: $i)$
each hadron looks like a pancake, Lorentz contracted in the
direction of the collision. Accordingly, the time it takes to parton $a$($b$),
within hadron $A$($B$),
to go through hadron $B$($A$) will be Lorentz contracted; $ii)$ the parton
interactions within each hadron are time dilated. Therefore, the time it takes
to a scattering parton to go through the other hadron is much less than the
interaction time between two partons in the same hadron. That is, parton
$a$($b$) sees hadron $B$($A$) as a frozen distribution of partons.
Besides, the momentum transfer $Q$ being large, parton $a$($b$) probes on
hadron $B$($A$) a region of transverse size $\sim 1/Q$, within which, unless
the hadron is densely packed, it will meet only parton $b$($a$).

Then the production process can be thought of as the
convolution of two contributions
\beq
\s = \sum_{ab} \int_{\zeta_A}^1 dx_A \int_{\zeta_B}^1 dx_B f_{a/A}(x_A,\mu^2)
f_{b/B}(x_B,\mu^2)
\hsig_{ab}\left(\zeta_A/x_A,\zeta_B/x_B,\a_s(\mu^2),Q^2/\mu^2\right),
\label{faca}
\eeq
where $a$ and $b$ are summed over the parton species and flavors;
$\zeta_A$, $\zeta_B$ and $\s$ are parametrized by the kinematic variables of
the produced hadrons or jets; $\hsig$ is the probability for the hard
scattering
to happen, which can be computed through perturbative quantum chromodynamics
(QCD) calculations; $f_{a/A}(x_A,\mu^2)$ is the probability
that parton $a$($b$) is found in hadron $A$($B$)
carrying a fraction $x_{A(B)}$ of the parent hadron momentum.
$f_{a/A}(x_A,\mu^2)$ is a universal quantity, i.e. it does not depend on
the particular hard scattering $\hsig$ considered,
it can not be computed in perturbation theory and must be derived from the
experimental data.

Eq.(\ref{faca}) expresses the factorization in the parton model, which is
valid in the region where $x$ is fixed and $Q^2 \ra \infty$. For finite,
and large, values of $Q$, the production cross section is expanded in
powers of $1/Q$; then eq.(\ref{faca}) may be shown to be valid for the leading
term of the expansion \cite{css}, called the leading twist from operator
product expansion terminology. The remainder of the expansion, which does
not usually factorize, is called the higher twist.

Then we proceed to calculate the partonic cross section $\hsig$, as an
expansion in the strong coupling constant $\alpha_s$
\beq
\hsig\left(\zeta/x,\a_s,Q^2/\mu^2\right) = \a_s^h
\left[\hsig^{(0)}\left(\zeta/x\right) + \sum_{n=1}^{\infty}
\a_s^n \hsig^{(n)}\left(\zeta/x,Q^2/\mu^2\right) \right],
\label{hard}
\eeq
where the power $h$ and the coefficients $\hsig^{(i)}$ depend on the production
process. In doing the calculation, we set
the parton masses and transverse momenta to zero, since
their contribution to eq.(\ref{faca}), of ${\cal O}(\Lambda_{QCD}/Q)$,
is in the higher twist.
Calculating the coefficients $\hsig^{(i)}$, with $i\ge 1$, we
encounter both ultraviolet and infrared divergences. The ultraviolet
divergences are due to the virtual radiative corrections, they are an artifact
of the perturbative expansion and are subtracted away by using a
renormalization prescription. This introduces in the calculation the dependence
on a renormalization scale $\mu$. Left over are then the infrared divergences,
divided into soft divergences, which appear when a parton momentum vanishes,
and collinear divergences, due to the collinear production of massless partons
in a vertex. After including all the virtual and real radiative corrections to
a given coefficient $\hsig^{(i)}$, the soft divergences
cancel out. The left-over collinear divergences are universal, i.e. are not
specific to the coefficient $\hsig^{(i)}$ we obtained them from. They are
the outcome of the collinear emissions from parton $a$($b$) on its evolution
toward the hard scattering. This evolution is space-like, hence during the
evolution the absolute value of the parton virtuality grows.
In particular the collinear emissions
describe the distribution of parton $c$($d$) of momentum fraction
$\zeta_{A(B)}$ and virtuality $\mu^2$, within parton $a$($b$) of momentum
fraction $x_{A(B)}$ and zero mass and transverse momentum.
Since the collinear parton evolution is universal we factorize it
into the parton density $f_{a/A}(x_A,\mu^2)$.

While both the parton density and the partonic cross section
depend on the renormalization/factorization scale\footnote{Even though they
are often chosen to be the same, the renormalization
and factorization scales are in principle unrelated (see sect. 14.3 of
ref.\cite{ster}).} $\mu$, their convolution
(\ref{faca}) does not, since the physical process does not depend on the
value of the virtuality $\mu^2$ we stop the collinear parton evolution at,
\beq
{d\s\over d\ln\mu^2} = 0.
\label{ren}
\eeq
Replacing into eq.(\ref{ren}) the factorization formula (\ref{faca}), we
obtain renormalization group equations for the coefficients $\hsig^{(i)}$ and
for the parton densities \cite{css}. The ones for the parton densities
\beq
{d f_{a/A}(x,\mu^2)\over d\ln\mu^2} = \sum_c \int_x^1 {d\xi\over\xi}
P_{ac}\left(\xi,\a_s(\mu^2)\right) f_{c/A}(x/\xi,\mu^2),
\label{rge}
\eeq
are called Altarelli-Parisi, or DGLAP, evolution equations \cite{dglap}, and
describe the distribution of parton $c$ within parton $a$,
as pictured in the paragraph above\footnote{An introduction to
eq.(\ref{rge}) and its relation to the renormalization group may be got
from the horse's mouth \cite{alt}.}. The
functions $P_{ac}$ are known as Altarelli-Parisi splitting functions and may
be computed in perturbative QCD as an expansion in $\a_s$
\beq
P_{ac}\left(\xi,\a_s\right) = \sum_{n=1}^{\infty}
\left({\a_s\over 2\pi}\right)^n P_{ac}^{(n-1)}(\xi).
\label{split}
\eeq
Eq.(\ref{rge}) resums the collinear logarithms in the evolution
at a given accuracy, determined by the splitting functions (\ref{split}).
Namely, the leading order (LO), or one-loop, splitting functions \cite{dglap}
yield the (all-order) leading logarithmic evolution in eq.(\ref{rge}), the
next-to-leading order (NLO), or two-loop, splitting functions \cite{nlo}
yield the next-to-leading logarithmic evolution in eq.(\ref{rge}), and so on.
Accordingly, the running of $\a_s$ in eq.(\ref{rge}) must be determined to the
corresponding loop-accuracy.

Eq.(\ref{ren}) states the indipendence of the physical cross section on the
factorization scale $\mu^2$, and it holds exactly if we know the full
expansions (\ref{hard}) and (\ref{split}). However, in a fixed-order expansion
of the factorization formula (\ref{faca}), say at ${\cal O}(\a_s^n)$,
eq.(\ref{ren}) holds only up to corrections of ${\cal O}(\a_s^{n+1})$.
Hence the more terms we know in the expansion of the partonic cross sections
(\ref{hard}) and of the splitting functions (\ref{split}), the less our
evaluation of eq.(\ref{faca}) depends on the unphysical scale $\mu^2$.

In inclusive jet production the partonic cross sections
$\hsig$ are known at best at NLO in $\a_s$. Namely,
at present 4-parton \cite{rke} and 5-parton \cite{dix} NLO matrix elements
are available. By combining the 4-parton NLO matrix elements
\cite{rke} with the 5-parton LO matrix elements \cite{exact}, NLO one-jet
\cite{EKS} and two-jet \cite{EKS2}-\cite{ggk} inclusive distributions have
been computed.
Besides reducing the dependence on the factorization scale $\mu^2$, they
allow one an analysis of the dependence of jet production on the jet-cone
size and of the parton distribution within the cone \cite{EKS}.
They appear to be in good agreement with the data on one-jet \cite{CDF}
and two-jet \cite{CDF2}-\cite{ssos} inclusive
distributions from the Fermilab Tevatron Collider.

\subsection{Large Rapidities and the BFKL Evolution}
\label{sec:onetwo}

In section~\ref{sec:oneone} we have outlined how the collinear factorization
applies to an
inclusive production process and we have quoted the main results of
the standard hadronic jet analysis. We have assumed that in the
production process there is only one hard scale $Q$, the momentum transfer,
of the order of the jet transverse energy; and that the hadron
center-of-mass energy $\sqrt{s}$ is of the same order as $Q$.
However, at the Fermilab Tevatron Collider and at future hadron
colliders the {\it semihard region} of the kinematic phase space, where
$\sqrt{s} \gg Q$, is accessible. If $\sqrt{\hat s}=\sqrt{x_Ax_Bs}$ is the
parton center-of-mass energy, then
\beq
\ln{s\over Q^2} \,=\, \ln{1\over x_A} \,+\, \ln{\hat s \over Q^2}
\,+\, \ln{1\over x_B}.
\label{grap}
\eeq
The logarithms, $\ln(1/x)$, appear in the evolution of the parton
densities; the logarithm, $\ln(\hat s /Q^2)$,
parametrizes the hard scattering $\hsig$.
In the semihard region the left-hand side of eq.(\ref{grap}) is large.
Then in the production process we may have large logarithms,
$\ln(1/x)$, or large logarithms, $\ln(\hat s /Q^2)$, related as we will
see to the occurrence of large rapidity intervals
between the produced jets. Thus, large non-collinear logarithms may have to be
resummed either in the splitting functions (\ref{split}) or
in partonic cross section (\ref{hard}).
A unified treatment of factorization which
takes into account the general case where both these logarithms are large,
and at the same time includes the collinear
factorization (\ref{faca}), is given in ref. \cite{cat}.

In these lectures, we deal with jet production in the semihard region,
at large momentum fractions $x$'s of the incoming partons \cite{MN}.
In this case there are no large
logarithms, $\ln(1/x)$, in the splitting functions (\ref{split}). Thus the
parton densities evolve according to the usual Altarelli-Parisi evolution
and the collinear factorization (\ref{faca}) is suitable to describe the
production process. However large logarithms, $\ln(\hat s /Q^2)$, may
appear in the partonic cross section (\ref{hard}), and we consider
the problem of resumming them.

Resummation techniques of large logarithms, $\ln(s/Q^2)$, date back
to the studies of the high-energy limit of quantum electrodynamics
(QED) \cite{gri}. It is easy to see using power-counting
arguments that at high center-of-mass energy $\sqrt{s}$ and fixed momentum
transfer $Q$, the leading contribution to a given scattering amplitude comes
from photon exchange in the crossed channel $t$. This contribution is
responsible for letting the total cross section approach constant values
at very high energies, in accordance with Pomeranchuk theorem \cite{pom}.
Let us suppose then that along the photon exchanged in the $t$ channel a
fermion pair is emitted. The ensuing ${\cal O}(\a^2)$ radiative corrections to
the scattering amplitude may contain large logarithms, $\ln(s/Q^2)$ \cite{gri}.
The leading powers of $\left[\a^2 \, \ln(s/Q^2)\right]$ may be resummed in the
perturbative expansion, and the resummed series yields
the total cross section for the amplitude with exchange of one
photon in the $t$ channel.

The same techniques then have been applied by Lipatov and collaborators to
non-abelian gauge theories \cite{lip}-\cite{BFKL2}, and in particular
to perturbative QCD \cite{bal}. Like in QED, in the limit of
high center-of-mass energy $\sqrt{\hat s}$ and fixed momentum
transfer $Q$ the leading contribution to a given scattering amplitude in a
physical gauge comes from gluon exchange in the $\hat t$ channel. However,
due to the gluon self-coupling interaction, the radiative corrections
which contain large logarithms, $\ln(\hat s /Q^2)$, appear at ${\cal O}(\a_s)$.
Therefore, they are roughly a factor $\a_s/\a^2$ stronger than the
corresponding corrections in QED, and may be of practical importance in
computations of high-energy scattering.

\section{Two-jet production at leading order}

In this section two-jet production at hadron colliders at the leading
order in $\a_s$ is discussed. We begin by defining the jet kinematic
variables by which the high-energy process will be described; the two-parton
kinematics as a function of the jet variables and the factorization
formula that connects the parton subprocess to jet production are introduced;
the parton dynamics is discussed and the relevant subprocess in the
large-rapidity limit is computed; two-jet production at LO at the Tevatron
and the LHC colliders is calculated using the kinematics and dynamics in the
exact form and in the large-$y$ approximation; finally, it is sketched how
two-jet production can be used to gain knowledge about the parton densities.

\subsection{Rapidity}
\label{sec:twoone}

Let us consider two inertial frames
with the longitudinal axes oriented along the
beam direction of the colliding particles, and in relative motion in the
beam direction. Then the motion of a particle, at rest in a frame, is
characterized in the other frame by the relation
\beq
\tanh{y} = {p_{_{||}}\over E},
\label{lor}
\eeq
where $E$, $p_{_{||}}$ and $y$ are respectively the particle energy,
longitudinal momentum and rapidity. Inverting eq.(\ref{lor}), we obtain the
definition of rapidity as
\beq
y = {1\over 2} \ln{E+\pcol \over E-\pcol}.
\label{rap}
\eeq
Eq.(\ref{lor}) states that
energy and longitudinal momentum scale like $\cosh{y}$ and $\sinh{y}$,
respectively. The constant of proportionality is given by the mass-shell
condition. Introducing the transverse mass,
$m_{\perp} = \sqrt{m^2 + \ptp^2}$, where $m$ is the particle mass and
$\ptp$ the absolute value of the momentum in the plane transverse to the
beam direction, the mass-shell condition reads
$E^2 = \pcol^2 + m_{\perp}^2$. Using it in
eq.(\ref{lor}), the particle energy and longitudinal momentum are expressed by
\bea
\pcol &=& m_{\perp} \sinh{y}; \label{mt} \\
E &=& m_{\perp} \cosh{y}. \nonumber
\eea
It is often convenient to introduce light-cone coordinates, by combining
energy and longitudinal momentum
\beq
p^{\pm}= E\pm\pcol\, ,\label{lc}
\eeq
then the scalar product of two vectors is
$p\cdot q = (p^+q^- + p^-q^+)/2 - p_{\perp}\cdot q_{\perp}$
and the non-vanishing components of the metric tensor are
\beq
2g_{+-} = 2g_{-+} = -g_{xx} = -g_{yy} = 1. \label{lcmetr}
\eeq
Using eq.(\ref{lc}), the particle 4-momentum is parametrized by mass,
transverse momentum and rapidity as
\beq
p = \left(m_{\perp}e^y, m_{\perp}e^{-y}; {\bf\ptp}\right),
\label{lcc}
\eeq
where ${\bf\ptp} = \left(\ptp\,\cos\phi, \ptp\,\sin\phi\right)$ and $\phi$ is
the azimuthal angle between the vector $\ptp$ and an arbitrary vector
in the transverse plane.
For massless particles we have from eq.(\ref{lor}) that
$\tanh{y}=\cos{\theta}$, with $\theta$ the angle between the directions of the
scattered particle and the beam. Then eq.(\ref{rap}) transforms to
\beq
y(m=0) = {1\over 2}\ln{1+\cos{\theta}\over 1-\cos{\theta}} =
- \ln\tan{\theta\over 2}. \label{prap}
\eeq
Eq.(\ref{prap}) defines the pseudo-rapidity, $\eta\equiv y(m=0)$. This, and
not the rapidity $y$, is the variable used in high-energy experiments, since
its determination requires just tracking the particle direction.
For massless particles, or in the case where $\ptp \gg m$,
the two definitions coincide. However, it is
worth reminding the difference, since it is the rapidity $y$ to define
the correct measure of the particle phase space
\beq
{d^3p\over 2E (2\pi)^3} = {dy\, d^2\ptp \over 4\pi (2\pi)^2},
\label{ph}
\eeq
and to transform additively under boosts in the beam direction.
Thus, the shape of a particle multiplicity distribution $dN/dy$
is boost invariant. To see the consequences of considering, on the contrary,
multiplicity spectra in pseudo-rapidity $dN/d\eta$, let us take a typical
hadron-hadron collision event, where the low transverse-momentum particles
from the underlying event fill up, rather uniformly, the phase space in $y$.
Now, consider eq.(\ref{mt}), and the analogous relation
for pseudo-rapidity $\sinh{\eta} = \pcol/\ptp$. Massive
particles with $\ptp \ll m$ will have $|\eta| > |y|$, hence they are
pushed away from the pseudo-rapidity central region $\eta \simeq 0$.
Then a fake excess of massless particles (photons) at $\eta \simeq 0$ is
found in the pseudo-rapidity distribution $dN/d\eta$ \cite{pump}.

\subsection{Elastic Scattering}
\label{sec:twotwo}

Let us consider the scattering of parton $a$ in hadron $A$ with parton
$b$ in hadron $B$, in the hadron center-of-mass frame, which we choose to be
the lab frame. In light-cone
coordinates, the 4-momenta of the incoming partons are
\bea
p_a &=& \left(\sqrt{s} x_A, 0; {\bf 0}\right), \label{in} \\
p_b &=& \left(0, \sqrt{s} x_B; {\bf 0}\right). \nn
\eea
The momenta of the outgoing partons, which at LO we identify with
the observed jets, are given according to the parametrization
(\ref{lcc}) by
\beq
k_i = \left(\kti e^{y_i}, \kti e^{-y_i}; {\bf\kti}\right), \label{out}
\eeq
with $i = a, b$. We introduce then the rapidity
boost $\bar y = (y_a+y_b)/2$, i.e. the rapidity of the parton center-of-mass
frame with respect to the lab frame, and the rapidity difference
$y = 2 y^* = y_a-y_b$, which is twice the rapidity $y^*$ of a parton in the
center-of-mass frame. Since in jet production the parton final states are
indistinguishable, without losing generality we assume that $y > 0$.

Transverse momentum conservation requires that $\kta = \ktb = \kt$, while
light-cone momentum conservation determines the momentum fractions $x$'s of the
incoming partons as a function of the jets kinematic variables
\begin{equation}
\begin{array}{ccccc}
x_A &=& \displaystyle
{\kt\over\sqrt{s}} e^{y_a} + {\kt\over\sqrt{s}} e^{y_b} &=& \displaystyle
2 {\kt\over\sqrt{s}} e^{\bar y} \cosh{y^*}, \\ \\
x_B &=& \displaystyle
{\kt\over\sqrt{s}} e^{-y_a} + {\kt\over\sqrt{s}} e^{-y_b} &=& \displaystyle
2 {\kt\over\sqrt{s}} e^{-\bar y} \cosh{y^*}. \label{exlo}
\end{array}
\end{equation}
Taking the ratio, we obtain $2 \bar y = \ln(x_A/x_B)$,
and in agreement with the definition of rapidity boost, $\bar y = 0$ when
the center-of-mass frame is at rest in the lab frame.
The Mandelstam invariants, which characterize the
parton scattering, are given in terms of the jet variables by
\bea
\hat s &=& 4 \kt^2 \cosh^2y^*, \nn \\
\hat t &=& -2 \kt^2 \cosh y^* e^{-y^*}, \label{mand} \\
\hat u &=& -2 \kt^2 \cosh y^* e^{y^*}. \nn
\eea
Note that the Mandelstam invariants do not depend on $\bar y $, since $\bar y$
does not belong to the center-of-mass frame kinematics. Using in
eq.(\ref{mand}) the relation between the Mandelstam invariants and the
scattering angle $\theta$ in the center-of-mass frame,
\bea
\hat t &=& -{\hat s \over 2} (1-\cos\theta), \label{scat} \\
\hat u &=& -{\hat s \over 2} (1+\cos\theta), \nn
\eea
we find $y^* = -\ln\tan(\theta/2)$, in agreement with the
definition (\ref{prap}) for massless partons.

\subsection{Two-jet production}
\label{sec:twotre}

Having described the parton kinematics, we have to establish how the parton
scattering is related to jet production at the hadron level. To do that, we
need a factorization formula, as outlined in the introduction. The
factorization formula for two-jet production cross section in terms of the
jet rapidities and transverse momenta is in general given by
\beq
{d\sigma\over d^2\kta d^2\ktb dy_a dy_b}\
=\ \sum_{ij} \int dx_Ax_B\,f_{i/A}(x_A,\mu^2)f_{j/B}(x_B,\mu^2)\,
{d\hat\sigma_{ij}\over d^2\kta d^2\ktb dy_a dy_b},
\label{jetfac}
\eeq
where $i$ and $j$ are summed over the parton species and flavors.
The parton cross section for elastic scattering is
\beq
d\hat\sigma_{ij} = {(2\pi)^4 \delta^4(p_a+p_b-k_a-k_b)\over 2\hat s}\,
{dy_a\, d^2\kta \over 4\pi\, (2\pi)^2}\, {dy_b\, d^2\ktb \over 4\pi\,
(2\pi)^2}\, |M_{ij}|^2, \label{born}
\eeq
where we have used the parametrization (\ref{ph}) for the phase space of
the outgoing partons, and where $|M_{ij}|^2$ are the squared matrix elements,
summed (averaged) over final (initial) parton helicities and colors\footnote{In
eq.(\ref{born}) we have not included the symmetry factor 1/2 for identical
final-state partons. We will keep that into account in the phase space with
$y > 0$ by summing only over the momentum configurations of different
final-state partons. This introduces no doublecounting, since everything
is symmetric under the exchange of the two jets.}.

Using eq.(\ref{exlo}) in the momentum-conserving $\delta$-function to fix the
parton momentum fractions, and replacing eq.(\ref{born}) into
eq.(\ref{jetfac}),
we obtain the factorization formula for two-jet production at LO
\beq
{d\sigma\over d\kt^2 dy_a dy_b}\
=\ \sum_{ij} x_A\, f_{i/A}(x_A,\mu^2)\, x_B\, f_{j/B}(x_B,\mu^2)\,
{d\hsig_{ij} \over d\hat t}, \label{jetlo}
\eeq
where the cross section for elastic scattering is
\beq
{d\hsig_{ij} \over d\hat t} = {|M_{ij}|^2 \over 16\pi\hat s^2}. \label{elas}
\eeq
\begin{figure}[hbt]
\vspace*{-14.0cm}
\hspace*{4.0cm}
\epsfxsize=15cm \epsfbox{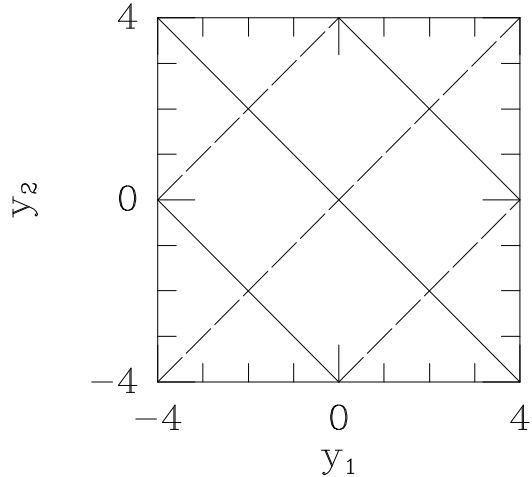}
\vspace*{-0.5cm}
\caption{Plane of the jet rapidities. The solid curves are lines of constant
${\bar y}$, the dashed curves are lines of constant $y$.}
\label{fig:ypl}
\end{figure}
As noted after eq.(\ref{mand}), the Mandelstam invariants,
and so $|M_{ij}|^2$, depend on the transverse momentum $\kt$ and the
rapidity difference $y$, but not on the boost $\bar y$; while the parton
momentum fractions $x$'s depend also on $\bar y$ (cf. eq.(\ref{exlo})). This
affects the choice of variables which characterize two-jet
production. To help visualize this, let us consider in Fig.~\ref{fig:ypl}
the plane of the jet rapidities. Then if the object of our study is the parton
scattering, it is convenient to either fix $\bar y$ (the solid lines of
Fig.~\ref{fig:ypl}) or integrate it out, since its
variation induces just a varying contribution from the parton densities.
If instead we are interested in the parton densities, then it is convenient
to fix $\kt$ and $y$ (the dashed lines of Fig.~\ref{fig:ypl}), thereby fixing
the contribution of the parton scattering to two-jet production.

\subsection{Parton Dynamics at Large Rapidities}
\label{sec:twofour}

In order to use the formula for two-jet production at LO (\ref{jetlo}), we
need
to compute the squared matrix elements $|M_{ij}|^2$. They have been evaluated
in ref.\cite{comb}, and are\footnote{The explicit calculation
for the subprocess $q\, g\ra q\, g$ can be found in sect. 7.2 of
ref.\cite{field}.}
\bea
|{\cal M}_{q\, q'\ra q\, q'}|^2 &=& |{\cal M}_{q\, \bar q'\ra q\, \bar q'}|^2 =
{4\over 9}{\hat s^2 + \hat u^2 \over \hat t^2}, \label{uno} \\
|{\cal M}_{q\, q\ra q\, q}|^2 &=& {4\over 9}\left({\hat s^2 + \hat u^2 \over
\hat t^2}+{\hat s^2 + \hat t^2 \over \hat u^2}\right) - {8\over 27}
{\hat s^2 \over \hat t \hat u}, \label{due} \\
|{\cal M}_{q\, \bar q\ra q\, \bar q}|^2 &=& {4\over 9}\left({\hat s^2 +
\hat u^2 \over \hat t^2}+{\hat t^2 + \hat u^2 \over \hat s^2}\right) -
{8\over 27} {\hat u^2 \over \hat t \hat s}, \label{tre}\\
|{\cal M}_{q\, \bar q\ra q'\, \bar q'}|^2 &=&
{4\over 9}{\hat t^2 + \hat u^2 \over \hat s^2}, \label{quattro}\\
|{\cal M}_{q\, \bar q\ra g\, g}|^2 &=& {32\over 27} {\hat t^2 + \hat u^2
\over \hat t \hat u} - {8\over 3} {\hat t^2 + \hat u^2 \over \hat s^2},
\label{cinque}\\
|{\cal M}_{g\, g\ra q\, \bar q}|^2 &=& {1\over 6} {\hat t^2 + \hat u^2
\over \hat t \hat u} - {3\over 8} {\hat t^2 + \hat u^2 \over \hat s^2},
\label{sei}\\
|{\cal M}_{q\, g\ra q\, g}|^2 &=& {\hat s^2 + \hat u^2 \over \hat t^2} -
{4\over 9} {\hat s^2 + \hat u^2 \over \hat s \hat u}, \label{sette}\\
|{\cal M}_{g\, g\ra g\, g}|^2 &=& {9\over 2}\left(3-{\hat t\hat u\over\hat s^2}
- {\hat s\hat u\over\hat t^2} - {\hat t\hat s\over\hat u^2}\right),
\label{otto}
\eea
where $q$ and $q'$ are quarks of different flavors, and
$|M_{ij}|^2 = g_s^4\, |{\cal M}_{ij}|^2$, with $g_s^2 = 4\pi\a_s$.
Now, we examine the dependence of the parton subprocesses
(\ref{uno}-\ref{otto}) on the rapidity interval $y = 2y^*$, and as a sample we
consider the subprocess $g\, g \ra g\, g$ (\ref{otto}).
Replacing in it the Mandelstam invariants (\ref{mand}), we obtain
\beq
|{\cal M}_{g\, g\ra g\, g}|^2 = {9\over 2}\, {\left(4\cosh^2y^* - 1\right)^3
\over 4\cosh^2y^*}, \label{ottob}
\eeq
then the corresponding parton cross section (\ref{elas}) is
\beq
{d\hsig_{gg\ra gg} \over d\hat t} = {9\over 2}\, \pi\,\a_s^2\, {1\over\kt^4}\,
{\left(4\cosh^2y^* - 1\right)^3 \over \left(4\cosh^2y^*\right)^3}.
\label{elasb}
\eeq
At small rapidities, eq.(\ref{elasb}) increases with $y$,
\beq
{d\hsig_{gg\ra gg} \over d\hat t} \sim 1 + y^{*2} - {1\over 3}y^{*4} +
{\cal O}(y^{*6}). \label{elasc}
\eeq

\begin{figure}[htb]
\vspace*{-5.0cm}
\hspace*{-0.0cm}
\epsfxsize=15cm \epsfbox{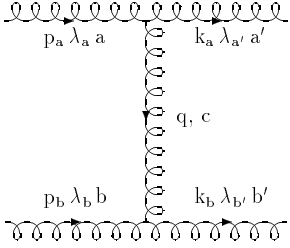}
\vspace*{-12.0cm}
\caption{$g\, g \ra g\, g$ scattering, with exchange of a gluon in the $\hat t$
channel. We label each external line with momentum, helicity and color.}
\label{fig:one}
\end{figure}

Then we consider the
large-$y$ limit. For $y \gg 1$, we find from eq.(\ref{mand}) that
\bea
\hat s &\simeq& -\hat u\,\simeq\, \kt^2 e^y, \label{ymand} \\
\hat t &\simeq& -\kt^2. \nn
\eea
Combining eq.(\ref{ymand}), the rapidity difference $y$ at large
rapidities becomes
\beq
y = \ln\left(-{\hat s \over \hat t}\right), \label{yrap}
\eeq
i.e. the large-$y$ limit is equivalent to the high-energy limit at fixed
momentum transfer. Four Feynman diagrams contribute to the subprocess
$g\, g \ra g\, g$: the
exchange of a gluon in the $\hat t$ channel (Fig.\ref{fig:one}); the exchange
of a gluon in the $\hat s$($\hat u$) channel, obtained from Fig.\ref{fig:one}
by crossing leg $k_a$ with leg $p_b$($k_b$); the four-gluon
coupling. Since $\hat s \simeq |\hat u| \gg |\hat t|$, only the diagram
with exchange of a gluon in the $\hat t$ channel contributes in the limit
$y \gg 1$, as long as we consider only physical polarizations for the external
gluons.

Inverting the metric tensor (\ref{lcmetr}) in order to let it act on the
helicity labels of a complete basis of (polarization-like) unit vectors
\beq
g^{\mu\nu} = v_{\lambda}^{\mu} g^{\lambda\lambda'} v_{\lambda'}^{\nu}
= 2(v_+^{\mu} v_-^{\nu} + v_-^{\mu} v_+^{\nu}) - v_{\perp}^{\mu}\cdot
v_{\perp}^{\nu},
\eeq
and using eq.(\ref{in}) in the light-cone vectors
\bea
v_+ &=& (1, 0, {\bf 0}) = {1\over x_A\sqrt{s}}\, p_a, \\
v_- &=& (0, 1, {\bf 0}) = {1\over x_B\sqrt{s}}\, p_b, \nn
\eea
we can rewrite the metric tensor as
\beq
g^{\mu\nu} = 2\,{p_a^{\mu}\, p_b^{\nu} + p_a^{\nu}\, p_b^{\mu} \over \hat s}
- \delta^{\mu\nu}_{\perp}, \label{metr}
\eeq
where $\delta^{\mu\nu}_{\perp}$ is a Kronecker delta over the transverse
components. Next, we compute the amplitude for the exchange
of a gluon in the $\hat t$ channel, to leading ${\cal O}(\hat s/|\hat t|)$,
and use eq.(\ref{metr}) in the gluon propagator.
Then the amplitude of Fig.\ref{fig:one} becomes
\bea
iM^{aa'bb'}_{\lambda_a\lambda_{a'}\lambda_b\lambda_{b'}} &\simeq&
g_s\, f^{aa'c}\, \left[g_{\mu_a\mu_{a'}} (p_a+k_a)_{\nu} + g_{\nu\mu_{a'}}
(-k_a+q)_{\mu_a} - g_{\mu_a\nu} (q+p_a)^{\mu_{a'}}\right] \nn\\
&\cdot& (-i) 2{p_a^{\rho}p_b^{\nu}\over \hs} {1\over\th} \nn\\
&\cdot& g_s\, f^{bb'c}\, \left[g_{\mu_b\mu_{b'}} (p_b+k_b)_{\rho} -
g_{\rho\mu_{b'}} (k_b+q)_{\mu_b} + g_{\mu_b\rho} (q-p_b)^{\mu_{b'}}\right]
\label{tchan} \\ &\cdot&
\epsilon_{\lambda_a}^{\mu_a}(p_a)\, \epsilon_{\lambda_b}^{\mu_b}(p_b)\,
\epsilon_{\lambda_{a'}}^{\mu_{a'}}(k_a)\,
\epsilon_{\lambda_{b'}}^{\mu_{b'}}(k_b), \nn \\ &\simeq&
-2i\, g_s^2\, f^{aa'c}\, g_{\mu_a\,\mu_{a'}}\,
{\hat s\over \hat t}\, f^{bb'c}\, g_{\mu_b\,\mu_{b'}}\,
\epsilon_{\lambda_a}^{\mu_a}(p_a)\, \epsilon_{\lambda_b}^{\mu_b}(p_b)\,
\epsilon_{\lambda_{a'}}^{\mu_{a'}}(k_a)\,
\epsilon_{\lambda_{b'}}^{\mu_{b'}}(k_b), \nn
\eea
where $f^{abc}$ are the $SU(N_c)$ structure constants, with $N_c=3$ the
number of colors, and $\epsilon_{\lambda}^{\mu}$ are
the gluon polarization vectors of helicity $\lambda$, and $q^2 = \th$.
Since in eq.(\ref{tchan}) we are considering physical polarizations,
$\epsilon_{\lambda}(p) \cdot p = 0$, the leading contribution
in $\hat s/|\hat t|$ comes from combining the helicity-conserving terms
in the 3-gluon vertices with one of the two light-cone polarization modes
in the gluon
propagator. Then, we square the amplitude and sum over helicities and
colors. We replace the sum over the gluon helicity states by
\beq
\sum_{\lambda}\, \epsilon_{\lambda}^{\mu}(p)\, \epsilon_{\lambda}^{\nu *}(p)\,
= - \left(g^{\mu\nu} - {n^{\mu}\, p^{\nu} + n^{\nu}\, p^{\mu}\over n\cdot p}
+ {n^2\, p^{\mu}\, p^{\nu}\over (n\cdot p)^2}\right), \label{axi}
\eeq
where $n$ is an arbitrary 4-vector. This is equivalent to use an axial gauge.
For example, a convenient choice for the sum over the helicity states of
gluon $p_a$ is to take $n\equiv p_b$. Then eq.(\ref{axi}) becomes
\beq
\sum_{\lambda_a}\, \epsilon_{\lambda_a}^{\mu_a}(p_a)\,
\epsilon_{\lambda_a}^{\nu_a *}(p_a)\, = - \left(g^{\mu_a\nu_a} -
{p_a^{\mu_a}\, p_b^{\nu_a} + p_a^{\nu_a}\, p_b^{\mu_a} \over p_a\cdot p_b}
\right) = \delta^{\mu_a\nu_a}_{\perp}, \label{tran}
\eeq
where the use of eq.(\ref{metr}) shows that only physical
polarization states are summed. Using eq.(\ref{tran}) in the amplitude
(\ref{tchan}) squared, we obtain for the sum over the helicity states of
gluons $p_a$ and $k_a$,
\beq
g^{\mu_a\mu_{a'}}\, g^{\nu_a\nu_{a'}}\, \left[\sum_{\lambda_a}\,
\epsilon_{\lambda_a}^{\mu_a}(p_a)\, \epsilon_{\lambda_a}^{\nu_a *}(p_a)\right]
\left[\sum_{\lambda_{a'}}\, \epsilon_{\lambda_{a'}}^{\mu_{a'}}(k_a)\,
\epsilon_{\lambda_{a'}}^{\nu_{a'}*}(k_a)\right]\, = 2
\left(1+{\cal O}\left({\th\over\hs}\right)\right), \label{hel}
\eeq
which shows that helicity is conserved at the jet-production
vertices, up to subleading terms.

The procedure we outlined in eq.(\ref{tchan}-\ref{hel}) is not gauge
invariant, since the terms we neglected in eq.(\ref{tchan}) are subleading
only if we consider physical polarizations, that is if we use a physical
gauge. In order for gauge invariance to hold, Ward identities must be
fulfilled, i.e. by replacing in the scattering amplitude one or more
of the physical polarizations with the longitudinal one,
$\epsilon_{\lambda}^{\mu}(p) \ra p^{\mu}$, the amplitude must
vanish\footnote{For a detailed discussion of gauge invariance and
Ward identities in non-abelian gauge theories, see
sect. 8.5 of ref.\cite{ster}, or sect. 7.2 of ref.\cite{field}.}.
It is easy to see that this is not the case in eq.(\ref{tchan}).
To obtain then a gauge-invariant amplitude,
one must consider the full expression of the
amplitude for the exchange of a gluon in the $\th$ channel, given in
eq.(\ref{tchan}), and include also the Feynman diagrams for
gluon exchange in the $\hat s$ or $\hat u$ channel and for four-gluon
coupling. Thus, amplitude (\ref{tchan}) assumes the form \cite{BFKL}
\beq
iM^{aa'bb'}_{\lambda_a\lambda_{a'}\lambda_b\lambda_{b'}} \simeq
-2i\, g_s^2\, f^{aa'c}\, \Gamma_{\mu_a\,\mu_{a'}}\,
{\hat s\over \hat t}\, f^{bb'c}\, \Gamma_{\mu_b\,\mu_{b'}}\,
\epsilon_{\lambda_a}^{\mu_a}(p_a)\, \epsilon_{\lambda_b}^{\mu_b}(p_b)\,
\epsilon_{\lambda_{a'}}^{\mu_{a'}}(k_a)\,
\epsilon_{\lambda_{b'}}^{\mu_{b'}}(k_b), \label{invborn}
\eeq
with
\beq
\Gamma^{\mu_a\,\mu_{a'}} = g^{\mu_a\,\mu_{a'}} -
{p_a^{\mu_{a'}} p_b^{\mu_a} + p_b^{\mu_{a'}} k_a^{\mu_a} \over p_a\cdot p_b} -
\th\, {p_b^{\mu_{a'}} p_b^{\mu_a}\over 2 (p_a\cdot p_b)^2},
\eeq
and the analogous expression for $\Gamma^{\mu_b\,\mu_{b'}}$, by exchanging the
labels $a$ and $b$.
It is straightforward to check that amplitude (\ref{invborn}) is gauge
invariant, up to subleading terms. Besides, we could
replace the sum over the gluon helicity states by
\beq
\sum_{\lambda} \epsilon_{\lambda}^{\mu}(p)
\epsilon_{\lambda}^{\nu *}(p) = -g^{\mu\nu}, \label{feyn}
\eeq
and obtain the right helicity counting,
\beq
\Gamma^{\mu_a\mu_{a'}}\, \Gamma^{\nu_a\nu_{a'}}\, \left[\sum_{\lambda_a}\,
\epsilon_{\lambda_a}^{\mu_a}(p_a)\, \epsilon_{\lambda_a}^{\nu_a *}(p_a)\right]
\left[\sum_{\lambda_{a'}}\, \epsilon_{\lambda_{a'}}^{\mu_{a'}}(k_a)\,
\epsilon_{\lambda_{a'}}^{\nu_{a'}*}(k_a)\right]\, = 2. \label{helb}
\eeq
However, considering only diagrams with
gluon exchange in the $\th$ channel, as we did in eq.(\ref{tchan}), is a
correct procedure as long as we work consistently in a physical gauge.

The sum over colors then yields
\beq
f^{aa'c}\, f^{a'ac'}\, f^{bb'c}\, f^{b'bc'}\, = C_A^2\, \left(N_c^2-1\right),
\label{col}
\eeq
where $C_A=N_c$ is the Casimir factor of the adjoint representation of $SU(3)$,
under which the gluons transform. Replacing eq.(\ref{hel}) and (\ref{col})
in the amplitude (\ref{tchan}) squared, and averaging over the
initial helicity and color states, we obtain
\beq
|M_{g\, g\ra g\, g}|^2 = {4 C_A^2 \over N_c^2-1}\, g_s^4\, {\hat s^2 \over
\hat t^2} = {9\over 2}\, g_s^4\, {\hat s^2 \over \hat t^2}, \label{yotto}
\eeq
which is in agreement with eq.(\ref{otto}), if there we take the Mandelstam
invariants in the large-rapidity limit (\ref{ymand}).

Then we examine the other subprocesses that contribute to two-jet production.
Replacing
eq.(\ref{ymand}) into eq.(\ref{uno}-\ref{sette}), we realize that
subprocesses (\ref{quattro}), (\ref{cinque}) and (\ref{sei}) do not give
a leading contribution in $(\hat s/|\hat t|)$, since they have no diagrams
with gluon exchange in the $\hat t$ channel. Subprocesses (\ref{uno}),
(\ref{due}), (\ref{tre}) and (\ref{sette}) yield, in the large-rapidity limit,
\bea
|M_{q\, q'\ra q\, q'}|^2 &=& |M_{q\, q\ra q\, q}|^2\, =\,
|M_{q\, \bar q\ra q\, \bar q}|^2\, =\,
{8\over 9}\,g_s^4\,{\hat s^2 \over \hat t^2}, \label{yuno} \\
|M_{q\, g\ra q\, g}|^2 &=& 2\,g_s^4\,{\hat s^2 \over \hat t^2}. \label{ysette}
\eea
Eq.(\ref{yuno}) and (\ref{ysette}) may be obtained by taking in
each subprocess the leading contribution to the diagram with gluon exchange
in the $\hat t$ channel. Taking the ratios of eq.(\ref{yotto}), (\ref{yuno}),
and (\ref{ysette}), we obtain
\beq
|M_{g\, g\ra g\, g}|^2\, =\,{9\over 4}\, |M_{q\, g\ra q\, g}|^2\, =\,
\left({9\over 4}\right)^2\, |M_{q\, q\ra q\, q}|^2, \label{ratio}
\eeq
where $9/4 = C_A/C_F$ is the relative color strength in the jet-production
vertices ($C_F\,=\,(N_c^2-1)/2N_c\,=\,4/3$ is the Casimir factor of the
fundamental representation of
$SU(3)$, under which the quarks transform). Thus it is enough to compute
one of the three subprocesses in eq.(\ref{ratio}) to describe the parton
dynamics in the large-rapidity limit. Having done it for the subprocess
$g\, g \ra g\, g$, we can include the contribution of the others in two-jet
production at LO (\ref{jetlo}) by using the effective parton density \cite{CM},
\beq
f_{eff}(x,\mu^2) = G(x,\mu^2) + {4\over 9}\sum_f
\left[Q_f(x,\mu^2) + \bar Q_f(x,\mu^2)\right], \label{effec}
\end{equation}
where the sum is over the quark flavors. We dropped the hadron label in
eq.(\ref{effec}), since the effective parton density is charge-conjugation
invariant. To complete the discussion of the parton dynamics at large
rapidities, we note by using eq.(\ref{elas}) and (\ref{yotto}), or taking
the large-$y$ limit of eq.(\ref{elasb}), that the parton cross section does
not fall off as the parton center-of-mass energy rises,
\beq
{d\hsig \over d\hat t} = {9\over 2}\, \pi\,\a_s^2\, {1\over\hat t^2}\, .
\label{yelas}
\eeq
This is in agreement with expectations of a constancy of the total
cross section at high energies \cite{pom}.

Also the kinematics simplify in the large-$y$ limit: the parton momentum
fractions (\ref{exlo}) become
\begin{eqnarray}
x_A^0 &=& {\kt\over\sqrt{s}} e^{y_a}, \label{largeyx} \\
x_B^0 &=& {\kt\over\sqrt{s}} e^{-y_b}, \nn
\end{eqnarray}
i.e. each parton momentum fraction is determined by the kinematic variables
of one jet only. Finally, we examine how the factorization formula
(\ref{jetlo}) tranforms in the large-$y$ limit.
The phase space for the production of two partons in eq.(\ref{born}) is
\beq
{\cal P}_2 = \int {dy_a\, d^2 k_{a\perp}\over 4\pi (2\pi)^2}\,
{dy_b\, d^2 k_{b\perp}\over 4\pi (2\pi)^2}\,
(2\pi)^4 \, \delta^4(p_a + p_b - k_a - k_b)\, .\label{twops}
\eeq
We can fix the rapidities using the light-cone momentum conservation
(\ref{exlo}) and the parametrizations (\ref{in}) and (\ref{out}),
\bea
&& 2\int dy_a\, dy_b\, \delta\left(\sqrt{s} x_A\, -\,
\kt e^{y_a} - \kt e^{y_b}\right)\,
\delta\left(\sqrt{s} x_B\, -\,
\kt e^{-y_a} - \kt e^{-y_b}\right)\, =\, \nn\\
&& {2\over \hs}\, {\cosh{(y_a-y_b)} + 1\over\sinh{(y_a-y_b)}}\, ,\label{jac}
\eea
where the overall factor 2 on the left-hand side comes from the Jacobian
in the change of variables (\ref{lc}).
Taking the large-$y$ limit in the right-hand side or
using the light-cone momentum conservation at large $y$ (\ref{largeyx}) in the
left-hand side, eq.(\ref{jac}) reduces to $2/\hs$. So the phase
space for the production of two partons (\ref{twops}) becomes, in the large-$y$
limit,
\beq
{\cal P}_2 = \int {1\over 2\hs}\, {d^2 k_{a\perp}\over (2\pi)^2}\,
{d^2 k_{b\perp}\over (2\pi)^2}\,
(2\pi)^2 \, \delta^2(k_{a\perp} + k_{b\perp})\, ,\label{mrtwops}
\eeq
and the cross section for elastic scattering (\ref{born}) reduces to
\beq
{d\hsig_{ij} \over d\kt^2} = {|M_{ij}|^2 \over 16\pi\hat s^2}\, ,\label{elxs}
\eeq
which coincides with eq.(\ref{elas}).
Using then eq.(\ref{effec}) the factorization formula
(\ref{jetlo}) becomes
\beq
{d\sigma\over d\kt^2 dy_a dy_b}\
=\ x^0_A\,f_{\rm eff}(x^0_A,\mu^2)\, x^0_B\,f_{\rm eff}(x^0_B,\mu^2)\,
{d\hat\sigma_{gg}\over d\kt^2}. \label{logen}
\end{equation}
Note that we could have obtained the factorization formula (\ref{logen})
from eq.(\ref{jetlo}) by
simply recalling (cf. eq.(\ref{ymand})) that in the large-$y$ limit
$\th\ra -\kt^2$. However, the procedure we have followed above is suitable
for generalizing eq.(\ref{logen}) to higher orders in $\a_s$.

\subsection{Phenomenology of two-jet production at leading order}
\label{sec:twofive}

We are now ready to perform a quantitative study of two-jet production, as a
function of the rapidity interval $y$ between the jets. First, we mention that
the dominance of $\th$-channel gluon exchange as $y$ increases, as described in
sect.~\ref{sec:twofour}, has been verified through a calculation of two-jet
production as a function of $y$ at LO and NLO in $\a_s$ \cite{EKS2}, which
has been found in good agreement with the corresponding data
from the Tevatron collider \cite{CDF2}, \cite{dzero}.

In this section we outline how to perform a LO calculation of two-jet
production,
collecting the information we have acquired in the previous sections, and
we examine the accuracy of the large-rapidity approximation at LO, at the
Tevatron proton-antiproton ($\sqrt{s} = 1.8$ TeV) and at the LHC proton-proton
($\sqrt{s} = 14$
TeV) colliders.  For an exact calculation we need the factorization
formula (\ref{jetlo}), with the parton cross section given by eq.(\ref{elas})
and (\ref{uno}-\ref{otto}). The Mandelstam invariants and the parton momentum
fractions are given in eq.(\ref{exlo}) and (\ref{mand}) as functions of the
jet kinematic variables. For the calculation in the large-$y$ approximation,
the factorization formula (\ref{logen}) is to be used, with the effective
parton density (\ref{effec}), and the parton cross section given by
eq.(\ref{elas}) and (\ref{yotto}). The corresponding
Mandelstam invariants and parton momentum fractions are given in
eq.(\ref{ymand}) and (\ref{largeyx}).

Since we are mainly interested in
understanding how the large-$y$ approximation works on the parton dynamics,
we fix the rapidity boost $\bar y = 0$
(cf. sect.~\ref{sec:twotwo} and \ref{sec:twotre}). Then we take
$\ptp \ge 20$~GeV, so that the parton densities are evaluated at
$x \gsim 2\cdot 10^{-2}$ at the Tevatron ($3\cdot 10^{-3}$ at the LHC).
The choice of the factorization scale $\mu^2$ is
arbitrary (cf. sect.~\ref{sec:oneone}); we fix its value at the jet
transverse energy, $\mu^2 = \ptp^2$. For the LO evolution of the
parton densities with $x$ and $\mu^2$, we take the CTEQ parton densities
\cite{CTEQ3}. Finally, we evaluate $\a_s(\mu^2)$, scaled from
$\a_s(m_Z^2)=0.12$, using the one-loop evolution with five flavors.

\begin{figure}[hbt]
\vspace*{-9.5cm}
\hspace*{-0.5cm}
\epsfxsize=15cm \epsfbox{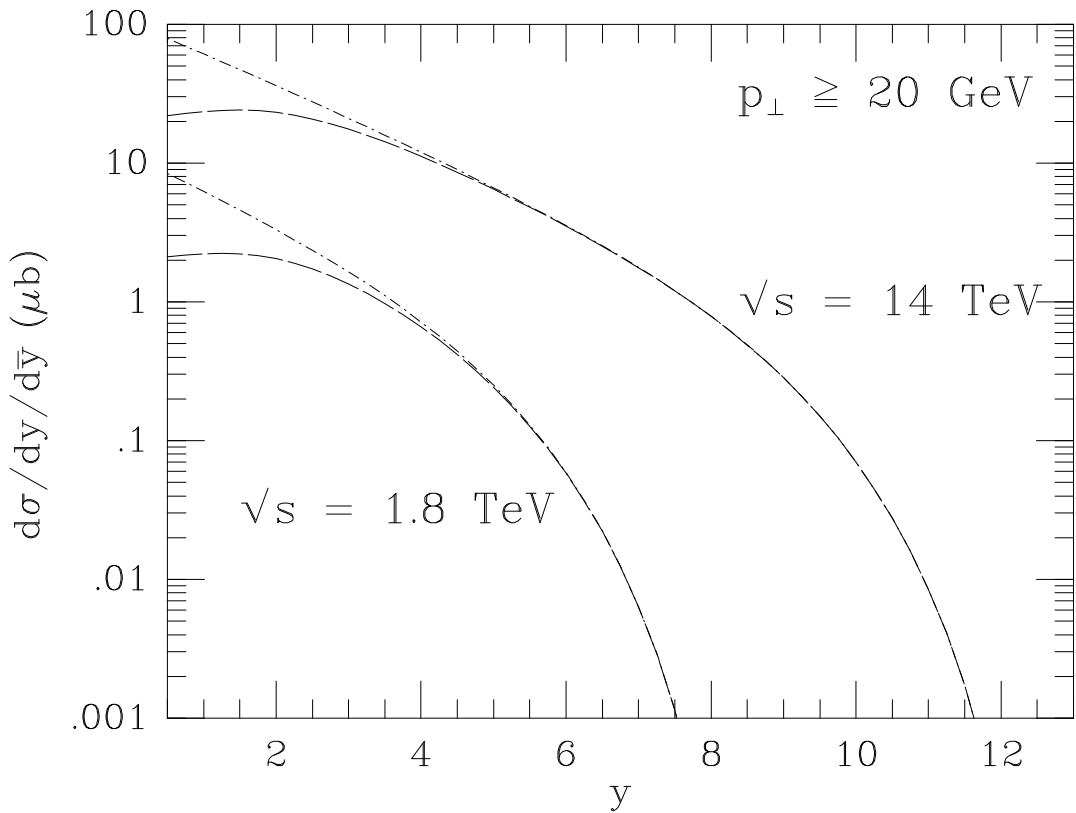}
\vspace*{-0.5cm}
\caption{Two-jet production at the Tevatron and LHC colliders, as a function
of the rapidity interval $y$, at $\bar y = 0$. The dashed and dot-dashed
lines are respectively the exact and large-$y$ LO cross sections.}
\label{fig:two}
\end{figure}

In Fig.~\ref{fig:two} we plot the two-jet production cross section
$d\s/dy\,d\bar y$ as a function of $y$, at the energies of the Tevatron
and LHC colliders. The dashed (dot-dashed) curves are obtained from the
exact (large-$y$) evaluation of the cross section. As expected the large-$y$
approximation improves as the size of the rapidity interval grows; at
$y \gsim 3$ the difference between the curves is about 20\%, thus, it is not
bigger than the theoretical uncertainty due to the choice of factorization
scale. The dot-dashed
curves of Fig.~\ref{fig:two} (the dashed ones, for large $y$) fall off
with $y$. This is due to the parton densities. Indeed at fixed $y$ the $x$'s
grow linearly with the jet transverse momentum (cf. eq.(\ref{exlo}) and
(\ref{largeyx})), thus at larger values of $y$ the integration
over $p_{\perp}$ starts at larger values of the $x$'s,
where the parton densities give a lesser contribution \cite{DDS}, \cite{DS},
\cite{stir}.
The behavior of the curves at small $y$ is different, though. From
eq.(\ref{yelas}) we know that the parton cross section in the large-$y$
approximation is constant as $y$ grows. This, together with the reduction in
phase space from the kinematics, yields the monotonic fall-off of the
dot-dashed curves. The exact parton cross section (\ref{elasb}), instead,
grows with $y$ at small rapidities (\ref{elasc}). This yields the initial rise
of the dashed curves, followed then by the parton-density suppression.

Next, we sketch how two-jet production may be used to obtain information
on the parton densities.
For $x \gsim 10^{-2}$ the quark densities are well known
from precision measurements of the structure function $F_2$ in deeply inelastic
scattering (DIS) \cite{bcdms}; from the target isospin dependence in Drell-Yan
scattering \cite{na}; from the $W^{\pm}$ asymmetry in hadron collisions
\cite{was}. The gluon density, though, is known only at $x \sim 0.4$ \cite{wa}.
For $x < 10^{-2}$ the gluon density is presently
extracted from the quark sea in measurements of $F_2$ in DIS \cite{hera}.
Also two-jet production may be used to determine the gluon density
\cite{ggk}, \cite{ssos}.
This can be achieved by considering two-jet production with the
jets on opposite sides (OS) on a plane in rapidity and azimuthal angle, the
lego plot, or with the jets on the same side (SS) of a lego plot.

In OS two-jet production we fix $\bar y = 0$ and let
$y = 2 y^*$ grow; the center-of-mass frame is at rest in the lab frame, and
the incoming partons carry the same fraction of the parent hadron energy,
$x_A = x_B$. From eq.(\ref{jetlo}), the cross section behaves like
\beq
\s_{OS} \sim \sum_{ij} f_{i/A}(x_A,\mu^2)\, f_{j/B}(x_A,\mu^2)\, \hsig_{OS}.
\label{os}
\eeq
At the Tevatron collider $x_A$ is typically $\gsim 10^{-2}$, and its lower
bound grows with $y$, as described in the paragraph above. Since at these
values of $x_A$ the quark density is well known \cite{bcdms}-\cite{was},
eq.(\ref{os}) may then be used to extract information on
the gluon density at the same values of $x_A$.

In SS two-jet production we fix $y = 0$ and let $\bar y = y^*$
grow; then one $x$, say $x_A$, grows and the other, $x_B$, decreases, i.e.
one of the incoming partons
is rather energetic and the other is wee, and the center-of-mass
frame is boosted by $\bar y$ in the direction of the energetic parton.
The wee parton is predominantly a gluon, since the gluon density dominates
at small $x$. Then the cross section behaves like
\beq
\s_{SS} \sim \sum_i f_{i/A}(x_A,\mu^2)\, g(x_B,\mu^2)\, \hsig_{SS}, \label{ss}
\eeq
with $x_B$ as small as $2\cdot 10^{-3}$ at the Tevatron collider. Since $x_A$
is now very large, the
parton density $f_{i/A}(x_A,\mu^2)$ is dominated by the valence quarks, and
its value is well known. So eq.(\ref{ss}) may be used to gain information on
the gluon density at small values of $x$.

\section{The perturbative QCD pomeron}

In this section higher-order corrections to two-parton production are
considered,
in the limit of a strong rapidity ordering of the final-state partons.
This kinematic regime yields the leading logarithmic contributions, of the type
$\ln(\hat s/|\hat t|)$, to the radiative corrections. Tree-level three-gluon
amplitudes, and the related production cross section, and one-loop elastic
amplitudes are computed in this limit. Multigluon amplitudes
are then introduced and used to evaluate the elastic
scattering amplitude with exchange of a two-gluon ladder in the $\th$ channel.
In doing this, the BFKL equation which describes the
gluon-ladder evolution in transverse momentum is derived\footnote{The BFKL
equation also describes the small-$x$ evolution of the gluon density, and in
that context it has been derived in ref.\cite{ciaf}, \cite{amuel}.}.
The solution is
found for color-octet \cite{BFKL}, and for color-singlet exchange \cite{BFKL2}.
The singlet solution is related through unitarity to the total
parton cross section, with exchange of a one-gluon ladder in the $\th$ channel
\cite{bal}.

\subsection{The {\sl n}-parton kinematics}
\label{sec:treone}

Let us assume that in the parton scattering $n+2$ partons are produced
(Fig.~\ref{fig:four}).
The 4-momenta of the incoming and outgoing partons are parametrized like
in eq.(\ref{in}) and (\ref{out}), with $i=0,...,n+1$. Momentum conservation
requires that
\bea
{\bf 0} &=& \sum_{i=0}^{n+1} {\bf\kti}, \nn \\
x_A &=& \sum_{i=0}^{n+1} {\kti\over\sqrt{s}} e^{y_i}, \label{nkin} \\
x_B &=& \sum_{i=0}^{n+1} {\kti\over\sqrt{s}} e^{-y_i}. \nn
\eea
Using eq.(\ref{nkin}), the Mandelstam invariants can be written as \cite{DDS2}
\begin{eqnarray}
\hat s &=& x_A x_B s = \sum_{i,j=0}^{n+1} k_{i\perp} k_{j\perp}
e^{y_i-y_j} \nonumber\\ \hat s_{ai} &=& -2 p_a\cdot k_i = -\sum_{j=0}^{n+1}
k_{i\perp} k_{j\perp} e^{-(y_i-y_j)} \label{inv}\\ \hat s_{bi} &=&
-2 p_b\cdot k_i = -\sum_{j=0}^{n+1} k_{i\perp} k_{j\perp} e^{y_i-y_j}
\nonumber\\ \hat s_{ij} &=& 2 k_i\cdot k_j = 2 k_{i\perp} k_{j\perp}
\left[\cosh (y_i-y_j) - \cos (\phi_i-\phi_j) \right]. \nonumber
\end{eqnarray}
Eq.(\ref{inv}) generalizes eq.(\ref{mand}). As in that case, we note that
the Mandelstam invariants depend only on rapidity differences, which again
reflects the boost invariance of the center-of-mass frame with respect to
the lab frame.

Next, we assume to work in the kinematic region where the outgoing partons
are strongly ordered in rapidity and have comparable transverse momentum,
of size $\kt$,
\beq
y_0 \gg y_1 \gg ...\gg y_{n+1};\qquad \kti\simeq\kt, \label{mreg}
\eeq
{}From eq.(\ref{inv}) and (\ref{mreg}), we obtain
\bea
\hat s &\simeq& k_{0\perp} k_{n+1\perp} e^{y_0-y_{n+1}}\, , \nn \\
\hat s_{ai} &\simeq& - k_{0\perp} k_{i\perp} e^{y_0-y_i}\, , \label{invb} \\
\hat s_{bi} &\simeq& - k_{i\perp} k_{n+1\perp} e^{y_i-y_{n+1}}\, , \nn \\
\hs_{ij} &\simeq& k_{i\perp} k_{j\perp} e^{|y_i-y_j|}\, ,\nn
\eea
then eq.(\ref{mreg}) can also be written as \cite{lip}, \cite{BFKL}
\bea
&& \hs \gg \hs_{ij} \gg k_{i\perp}^2, \label{mregge} \\
&& \prod_{i=0}^n \hs_{i,i+1}\, \simeq\, \hs \prod_{i=1}^n k_{i\perp}^2 \, . \nn
\eea
The kinematic region where eq.(\ref{mreg}), or eq.(\ref{mregge}), is valid
is called the {\sl multiregge kinematics} \cite{gri}, \cite{lip}-\cite{bal}.
The parton momentum fractions (\ref{nkin}) assume here the simple form
\bea
x_A^0 &\simeq& {k_{0\perp}\over\sqrt{s}} e^{y_0}, \label{ykin} \\
x_B^0 &\simeq& {k_{n+1\perp}\over\sqrt{s}} e^{-y_{n+1}}. \nn
\eea
which trivially generalizes eq.(\ref{largeyx}). Then we introduce
the vector $q_1 = p_a - k_0$, which labels the first momentum exchanged in
the $\th$ channel (Fig.~\ref{fig:four}). Using eq.(\ref{in}), (\ref{out})
and (\ref{nkin}), and keeping only the leading terms,
\beq
q_1 \simeq \left(k_{1\perp} e^{y_1},\,
-k_{0\perp} e^{-y_0},\, -{\bf k_{0\perp}}\right). \label{tm}
\eeq
Squaring $q_1$ and retaining only the leading term,
\beq
q_1^2 \simeq -k_{0\perp}^2\, =\, -q_{1\perp}^2, \label{tdeg}
\eeq
i.e. only the transverse degrees of freedom are relevant in the momentum
transfer $q_1$. The analysis can be repeated for the second momentum
exchanged in the $\th$ channel, $q_2 = q_1 - k_1$, with the same conclusions,
and so on for the following vectors $q_i$, i.e. in multiregge kinematics
only the transverse components are relevant to parametrize the propagators of
the gluons exchanged in the $\th$ channel,
\beq
\th_i = q_i^2 \simeq -q_{i\perp}^2. \label{tman}
\eeq

\subsection{Three-parton production in multiregge kinematics}
\label{sec:tretwo}

\begin{figure}[hbt]
\vspace*{-3.5cm}
\hspace*{-1.5cm}
\epsfxsize=15cm \epsfbox{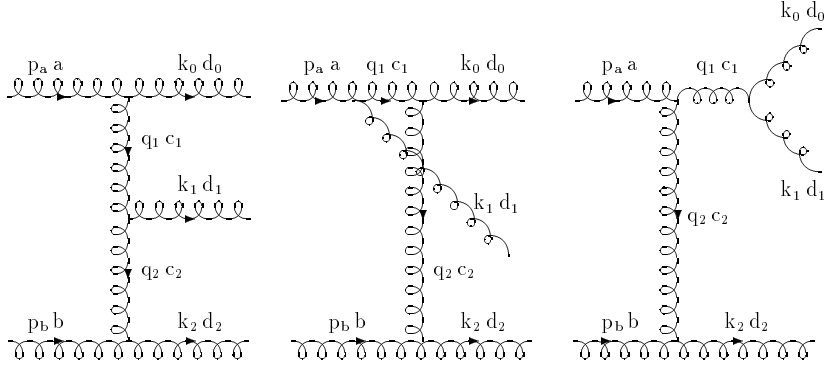}
\vspace*{-12.5cm}
\caption{$2\ra 3$-gluons Feynman diagrams, with gluon $k_1$ emitted
$a)$ from the gluon exchanged in the $\hat t$ channel; $b)$ in the
initial-state bremsstrahlung from the upper line; $c)$ in the final-state
bremsstrahlung from the upper line. We have neglected the helicity labels,
and we have omitted the diagrams with bremsstrahlung emission from the
lower line.}
\label{fig:tre}
\end{figure}

First, we compute the simplest amplitude in the multiregge kinematics,
the $2\ra 3$-gluons amplitude (Fig.~\ref{fig:tre}). As discussed in
sect.~\ref{sec:twofour}, we may consider just diagrams with gluon exchange
in the $\th$ channel, as long as we work in a physical gauge.
We consider first the
diagram of Fig.~\ref{fig:tre}a. Using the decomposition (\ref{metr}) for the
gluon propagators in the $\th$ channel, the corresponding amplitude is
\bea
iM(t)^{abd_0d_1d_2}_{\mu_a\mu_b\mu_0\mu_1\mu_2} &\simeq&
\left(g_s\, f^{ad_0c_1}\, g_{\mu_a\,\mu_0} \right)\, \hs (-i)\,
{2\over \hs\th_1} p_a^{\nu_1} \nn \\ &\cdot& g_s\, f^{c_1c_2d_1}\,
\left[g^{\nu_1\nu_2} (q_1+q_2)^{\mu_1} + g^{\mu_1\nu_2}
(-q_2+p_1)^{\nu_1} - g^{\nu_1\mu_1} (p_1+q_1)^{\nu_2}\right] \nn
\\ &\cdot& (-i)\, p_b^{\nu_2}\, {2\over \hs\th_2} \hs
\left(g_s\, f^{bd_2c_2}\, g_{\mu_b\,\mu_2} \right), \label{trega}
\eea
where we have implied
physical polarizations for the external gluons. Then, we introduce the
vector $\hat C^{\mu}(q_1,q_2)$, which
describes the contractions on the $q_1q_2k_1$-gluons vertex,
\bea
\hat C^{\mu_1}(q_1,q_2) &\equiv& {2\over \hs}\, p_a^{\nu_1}
\left[g^{\nu_1\nu_2} (q_1+q_2)^{\mu_1} + g^{\mu_1\nu_2}
(-q_2+p_1)^{\nu_1} - g^{\nu_1\mu_1} (p_1+q_1)^{\nu_2}\right] p_b^{\nu_2} \nn\\
&\simeq& (q_1+q_2)^{\mu_1}_{\perp} - {\hs_{a1}\over\hs} p_b^{\mu_1}
+ {\hs_{b1}\over\hs} p_a^{\mu_1}\, ,\label{cver}
\eea
where we have used the approximate form (\ref{invb}) of the Mandelstam
invariants, and the decomposition (\ref{metr}) of $(q_1+q_2)^{\mu}$,
\beq
(q_1+q_2)^{\mu} \simeq (q_1+q_2)^{\mu}_{\perp} + {\hs_{a1}\over\hs}
p_b^{\mu} - {\hs_{b1}\over\hs} p_a^{\mu}\, ,\label{deco}
\eeq
with $q_{\perp}^{\mu}=(0,0,{\bf q}_{\perp})$, such that $q_{\perp}\cdot
q_{\perp} = - q_{\perp}^2$. Replacing eq.(\ref{cver}) into eq.(\ref{trega}),
the amplitude for the emission of gluon $k_1$ along the $\th$-channel gluon is
\bea
iM(t)^{abd_0d_1d_2}_{\mu_a\mu_b\mu_0\mu_1\mu_2} &\simeq& 2i \hs
\left(i g_s\, f^{ad_0c_1}\, g_{\mu_a\,\mu_0} \right)\, {1\over \th_1} \nn\\
&\cdot& \left(i g_s\, f^{c_1d_1c_2} \hat C^{\mu_1}(q_1,q_2)\right)\,
{1\over \th_2} \label{tregb} \\
&\cdot& \left(i g_s\, f^{bd_2c_2}\, g_{\mu_b\,\mu_2} \right).\nn
\eea
The calculation of the amplitudes for the initial and final bremsstrahlung
emissions (Fig.~\ref{fig:tre}b,c) from the upper line yields respectively
\bea
iM(i)^{abd_0d_1d_2}_{\mu_a\mu_b\mu_0\mu_1\mu_2} &\simeq& -2 \hs\, g_s^3\,
f^{ac_1d_1}\, f^{c_1d_0c_2}\, f^{bd_2c_2}\, {2 p_a^{\mu_1} \over \hs_{a1}
\th_2}\, g_{\mu_a\,\mu_0}\, g_{\mu_b\,\mu_2}\, ,\label{brema}\\
iM(f)^{abd_0d_1d_2}_{\mu_a\mu_b\mu_0\mu_1\mu_2} &\simeq& 2 \hs\, g_s^3\,
f^{c_1d_0d_1}\, f^{ac_1c_2}\, f^{bd_2c_2}\, {2 p_a^{\mu_1} \over \hs_{a1}
\th_2}\, g_{\mu_a\,\mu_0}\, g_{\mu_b\,\mu_2}\, .\nn
\eea
Reordering the color with the Jacobi identities
\beq
f^{ac_1c_2}\, f^{d_0d_1c_1}\, +\, f^{d_0c_1c_2}\, f^{d_1ac_1}\, +\,
f^{d_1c_1c_2}\, f^{ad_0c_1}\, = \, 0\, , \label{jacob}
\eeq
the sum of the amplitudes for the bremsstrahlung emission from the upper line
gives
\bea
i \left(M(i)\,+\, M(f)\right)^{abd_0d_1d_2}_{\mu_a\mu_b\mu_0\mu_1\mu_2}
&\simeq& 2i \hs
\left(i g_s\, f^{ad_0c_1}\, g_{\mu_a\,\mu_0} \right)\, {1\over \th_1} \nn\\
&\cdot& \left(i g_s\, f^{c_1d_1c_2} {2\th_1\over\hs_{a1}}\, p_a^{\mu_1}\right)
\,{1\over \th_2} \label{bremb} \\
&\cdot& \left(i g_s\, f^{bd_2c_2}\, g_{\mu_b\,\mu_2} \right).\nn
\eea
Adding the amplitudes for the emission of the gluon from the $\th$
channel (\ref{tregb}), the bremsstrahlung emission from the upper line
(\ref{bremb}), and the bremsstrahlung emission from the lower line (whose
calculation is similar to the one of eq.(\ref{brema})), we obtain the
$2\ra 3$-gluons amplitude in the multiregge kinematics,
\bea
iM^{abd_0d_1d_2}_{\mu_a\mu_b\mu_0\mu_1\mu_2} &\simeq& 2i \hs
\left(i g_s\, f^{ad_0c_1}\, g_{\mu_a\,\mu_0} \right)\, {1\over \th_1} \nn\\
&\cdot& \left(i g_s\, f^{c_1d_1c_2}\, C^{\mu_1}(q_1,q_2)\right)\,
{1\over \th_2} \label{treg} \\
&\cdot& \left(i g_s\, f^{bd_2c_2}\, g_{\mu_b\,\mu_2} \right),\nn
\eea
where
\beq
C^{\mu_1}(q_1,q_2) \simeq \left[(q_1+q_2)^{\mu_1}_{\perp}\, -\,
\left({\hs_{a1}\over\hs}\,+\,2{\th_2\over\hs_{b1}}\right) p_b^{\mu_1}\,
+ \left({\hs_{b1}\over\hs}\,+\,2{\th_1\over\hs_{a1}}\right) p_a^{\mu_1}\right],
\label{lipver}
\eeq
is the non-local effective Lipatov vertex \cite{lip}, which summarizes the
insertion of the third gluon along the $\th$ channel, and as a bremsstrahlung
gluon. The Lipatov vertex is gauge invariant, indeed replacing the physical
polarization with the longitudinal one, the Ward identity
\beq
C^{\mu}(q_1,q_2)\, (k_1)_{\mu}\,=\, 0\, , \label{gauge}
\eeq
holds. Now we want to show that the amplitude (\ref{treg}) yields indeed
$\ln{(\hs/|\th|)}$ corrections to the Born-level two-parton production
considered
in sect.~\ref{sec:twofour}. We square the amplitude (\ref{treg}) and sum over
helicities and colors. The sum over the helicity of gluon $k_1$ may be
performed using eq.(\ref{feyn}) since the Lipatov vertex is gauge invariant
(\ref{gauge}),
\beq
C^{\mu}(q_1,q_2)\cdot C_{\mu}(q_1,q_2)\, =\, 4\, {q_{1\perp}^2 q_{2\perp}^2
\over k_{1\perp}^2}\, ,\label{csq}
\eeq
where we have used the kinematic constraint $\hs_{a1}\hs_{b1}\, =\,
k_{1\perp}^2 \hs$ from eq.(\ref{invb}). The sum over the helicities of
gluons $k_0$ and $k_2$ is performed using eq.(\ref{tran}). By using then
eq.(\ref{tdeg}) and (\ref{tman}) the square of amplitude (\ref{treg}),
summed (averaged) over final (initial) helicities and colors, is
\beq
|M_{gg\ra ggg}|^2\, =\, {16 N_c^3 g_s^6\over N_c^2-1}\, {\hs^2\over
k_{0\perp}^2 k_{1\perp}^2 k_{2\perp}^2}\, .\label{trgsq}
\eeq
Eq.(\ref{trgsq}) may be also obtained by taking the exact (square) matrix
elements for three-gluon production in gluon-gluon scattering \cite{exact},
\cite{pt},
\beq
|M_{gg\ra ggg}|^2\, =\,4\,(\pi\alpha_sN_c)^3 \, \sum_{i>j} \, \hs_{ij}^4 \,
\sum_{[a,0,1,2,b]'} \, {1 \over \hs_{a0}\hs_{01}\hs_{12}\hs_{2b}\hs_{ab}},
\label{parke}
\end{equation}
with i,j~=~a,0,1,2,b, and with the second sum over the noncyclic permutations
of the set [a,0,1,2,b]. In the strong rapidity ordering (\ref{mreg}) and
(\ref{invb}), eq.(\ref{parke}) reduces to eq.(\ref{trgsq}) \cite{vdd}.

Next, we must examine how the phase space for three-parton production,
\beq
{\cal P}_3 \equiv \prod_{i=0}^2 \int {dy_i\, d^2 k_{i\perp}\over 4\pi
(2\pi)^2}\, (2\pi)^4 \, \delta^4(p_a + p_b - \sum_{i=0}^2 k_i) \label{thpps}
\eeq
transforms in multiregge kinematics.
As we have seen in eq.(\ref{ykin}), light-cone momentum conservation at
large $y$ (\ref{largeyx}) simply generalizes to the multiregge kinematics.
Accordingly, eq.(\ref{jac}) also applies to three partons, and using it in
eq.(\ref{thpps}) the phase space becomes
\beq
{\cal P}_3 = \int {1\over 2\hs}\, {d^2 k_{0\perp}\over (2\pi)^2}\,
\left({dy_1\, d^2 k_{1\perp}\over 4\pi (2\pi)^2}\right)\,
{d^2 k_{2\perp}\over (2\pi)^2}\, (2\pi)^2 \, \delta^2\left(\sum_{i=0}^2
k_{i\perp}^2\right)\, , \label{mrtps}
\eeq
which straightforwardly generalizes eq.(\ref{mrtwops}). The cross section
for three-gluon production is then
\beq
\hsig_{gg\ra ggg}\, =\, {1\over 2\hs}\, {\cal P}_3\, |M_{gg\ra ggg}|^2\,
,\label{tpxs}
\eeq
and using eq.(\ref{trgsq}) and (\ref{mrtps}) and the transverse-momentum
conservation it becomes
\beq
{d\hsig_{gg\ra ggg}\over dk_{0\perp}^2 dk_{2\perp}^2 d\phi}\, =\, {N_c^3 \a_s^3
\over 4\pi}\, {y\over k_{0\perp}^2 k_{2\perp}^2 (k_{0\perp}^2 + k_{2\perp}^2
+ 2k_{0\perp}k_{2\perp}\cos{\phi})}\, ,\label{thpxs}
\eeq
where we have performed the integration of the rapidity of gluon $k_1$ over
the range $y=y_0-y_2$, and where
$\phi$ is the azimuthal angle between the transverse momenta $k_{0\perp}$
and $k_{2\perp}$. Since $y=\ln{(\hs/\kt^2)}$, eq.(\ref{thpxs}) shows the
logarithmic enhancement of the three-parton production with respect to the
two-parton production considered in sect.\ref{sec:twofour}. It also shows
that the intermediate gluon is produced with equal probability over the allowed
range in rapidity, i.e. eq.(\ref{thpxs}) is insensitive to the position in
rapidity of the intermediate gluon. This approximation has important
phenomenological consequences, and will be modified by
subleading corrections, as we will see in sect.~\ref{sec:fourtre}.

\begin{figure}[hbt]
\vspace*{-5.cm}
\hspace*{-0.5cm}
\epsfxsize=15cm \epsfbox{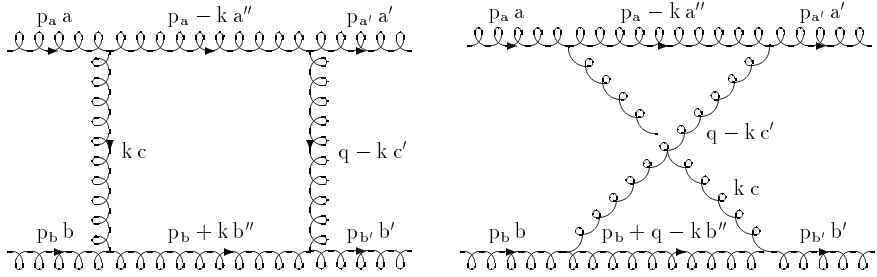}
\vspace*{-12.5cm}
\caption{One-loop corrections to the elastic amplitude $g g\ra g g$, with
$\hat t$-channel exchange of two gluons $a)$ in the $\hs$-channel physical
region and $b)$ in the $\hat u$-channel physical region.}
\label{fig:virt}
\end{figure}

\subsection{Virtual radiative corrections}
\label{sec:trenew}

Next, we consider the one-loop corrections to the subprocess $g g\ra g g$.
In the large-$y$ limit the diagrams with $\th$-channel exchange of
two gluons in the $\hs$-channel physical region
(Fig.\ref{fig:virt}a) and in the $\hat u$-channel physical region
(Fig.\ref{fig:virt}b) contribute. Let us consider the amplitude of
Fig.\ref{fig:virt}a and parametrize the momentum transfer $q^2=\th$ by the
vector $q\simeq (0,0,{\bf q}_{\perp})$, where following the discussion at the
end of sect.~3.1 only the transverse degrees of freedom are considered.
Using the decomposition (\ref{metr}) for the gluon propagators in the $\th$
channel, the leading contribution to the amplitude of Fig.\ref{fig:virt}a is
\beq
iM^{aa'bb'}_{\mu_a\mu_{a'}\mu_b\mu_{b'}} \simeq 2\hs^3\, g_s^4\, f^{aa''c}
f^{a''a'c'} f^{bb''c} f^{b''b'c'}\, g_{\mu_a\mu_{a'}} g_{\mu_b\mu_{b'}}\, I
,\label{vira}
\eeq
with $I$ the integral over the gluon propagators,
\bea
I &=& \int {d\alpha d\beta d^2k_{\perp}\over (2\pi)^4}\,
{1\over \alpha\beta\hs - \kt^2 + i\epsilon}\,
{1\over -(1-\alpha)\beta\hs - \kt^2 + i\epsilon}\nn\\
& & {1\over \alpha(1+\beta)\hs - \kt^2 + i\epsilon}\,
{1\over \alpha\beta\hs - (q-k)_{\perp}^2 + i\epsilon}\, ,\label{viri}
\eea
where we have decomposed the momentum $k$ on the light cone as
\beq
k^{\mu} = \alpha p_a^{\mu} + \beta p_b^{\mu} + k_{\perp}^{\mu}, \quad
{\rm with} \quad 0 < \alpha,\beta < 1\, ,\label{virs}
\eeq
and $d^4k = (\hs/2) d\alpha d\beta d^2k_{\perp}$ the phase-space measure.
Eq.(\ref{virs}) is called Sudakov decomposition. In eq.(\ref{viri})
all the propagators, but the second, have a pole in the
lower complex half-plane of $\beta$, so the integral over $\beta$ may be
easily performed in the upper half-plane, and we obtain
\beq
I \simeq -{i\over\hs}\,\int {d\alpha d^2k_{\perp}\over (2\pi)^3}\, {1\over
\alpha\hs-\kt^2} {1\over\kt^2} {1\over (q-k)_{\perp}^2}\, ,\label{virj}
\eeq
The integral of $\alpha$ is then logarithmic over the range
$(\kt^2/\hs) < \alpha < 1$. Performing it and substituting eq.(\ref{virj})
back into eq.(\ref{vira}), the amplitude of Fig.\ref{fig:virt}a becomes
\beq
iM^{aa'bb'}_{\mu_a\mu_{a'}\mu_b\mu_{b'}} \simeq -i {16\pi\a_s\over N_c}\,
{\hs\over\th}\, \ln{\hs\over -\th}\, \a(\th)\, f^{aa''c} f^{a''a'c'}
f^{bb''c} f^{b''b'c'}\, g_{\mu_a\mu_{a'}} g_{\mu_b\mu_{b'}}\,
,\label{virb}
\eeq
where the adimensional function $\a(\th)$ collects the loop
transverse-momentum integrations,
\beq
\alpha(\th) = \a_s\, N_c\, \th \int {d^2k_{\perp}\over (2\pi)^2}\, {1\over
\kt^2 (q-k)_{\perp}^2}\, .\label{allv}
\eeq
The integral in $\alpha(\th)$ may be evaluated by introducing an infrared
cutoff $\mu$
\beq
\a(\th) \simeq - {\a_s N_c\over 4\pi} \ln{q_{\perp}^2\over \mu^2}\, ,
\label{sudak}
\eeq
which shows that the amplitude (\ref{virb}) is doubly logarithmic divergent.
The amplitude of Fig.\ref{fig:virt}b is obtained by crossing the channels
$\hs$ and $\hat u$ in eq.(\ref{virb}). Using then
$\hat u = -\hs -\th$, and summing the diagrams of Fig.\ref{fig:virt}, the
amplitude for the exchange of two gluons in the $\th$ channel is,
to leading order in $\hs/\th$,
\bea
iM^{aa'bb'}_{\mu_a\mu_{a'}\mu_b\mu_{b'}} &\simeq& -i {16\pi\a_s\over N_c}\,
{\hs\over\th}\, g_{\mu_a\mu_{a'}} g_{\mu_b\mu_{b'}}\, \a(\th)\, \label{virc}\\
& & f^{aa''c} f^{a''a'c'} \left[\ln{\hs\over -\th} f^{bb''c} f^{b''b'c'} -
\left(\ln{\hs\over -\th} +i\pi\right) f^{bb''c'} f^{b''b'c}\right]\, .\nn
\eea
The color dependence may be decomposed as a sum over the $SU(3)$
representations in the Kronecker product of the adjoint representations,
$(\un{8}\otimes\un{8})$, for the two gluons exchanged in the $\th$ channel,
\beq
iM^{aba'b'}_{\mu_a\mu_b\mu_{a'}\mu_{b'}}\, =\, i\, g^{\mu_a\mu_{a'}}\,
g^{\mu_b\mu_{b'}}\, \sum_T P^{aa'}_{bb'}(T) A^T(\hs, \th), \label{elasgen}
\eeq
where we have also singled out the helicity dependence. $A^T(\hs, \th)$ are
then scalar amplitudes, and $P^{aa'}_{bb'}(T)$ are color projectors
\beq
P^{aa'}_{bb'}(T)\, P^{bb'}_{cc'}(T')\, =\, P^{aa'}_{cc'}(T) \delta_{TT'}\, .
\label{proj}
\eeq
For the exchange in the $\th$ channel of a representation in the
antisymmetric part of the product, $(\un{8}\otimes\un{8})_A=\un{8}_A\oplus
\un{10}\oplus\overline{\un{10}}$,
or of a singlet, which is in the symmetric part of the product,
$(\un{8}\otimes\un{8})_S=\un{1}\oplus\un{8}_S\oplus\un{27}$, the color
projectors are \cite{lipat}, \cite{mac}
\bea
P^{aa'}_{bb'}(\un{1}) &=& {1\over N_c^2-1} \delta^{aa'} \delta_{bb'}\, ,\nn\\
P^{aa'}_{bb'}(\un{8}_A) &=& {1\over N_c}\, f^{aca'} f^{bcb'}\,
,\label{octsin}\\
P^{aa'}_{bb'}(\un{10}\oplus\overline{\un{10}}) &=& {1\over 2}(\delta_{ab}
\delta_{a'b'} - \delta_{ab'}\delta_{a'b}) - {1\over N_c} f^{aca'} f^{bcb'}\,
.\nn
\eea
The projectors (\ref{octsin}) are normalized in order to satisfy the
projection rule (\ref{proj}). But for the singlet which is going
to be of later use, we do not report explicitly the projectors of
$(\un{8}\otimes\un{8})_S$, since the leading contribution of it to
eq.(\ref{virc}) cancels out because of the projectors parity under
$\hs \leftrightarrow \hat u$ crossing,
\beq
P^{aa'}_{b'b}(T)\, =\, (-1)^T\, P^{aa'}_{bb'}(T)\, ,\label{parit}
\eeq
with
\beq
(-1)^T = \left\{ \begin{array}{ll} -1 & \mbox{for $(\un{8}\otimes\un{8})_A$},\\
+1 & \mbox{for $(\un{8}\otimes\un{8})_S$}, \end{array} \right. \label{cp}
\eeq
The contraction of the structure constants in
eq.(\ref{virc}) with $P^{aa'}_{bb'}(\un{10}\oplus\overline{\un{10}})$ from
eq.(\ref{octsin}) cancels out exactly, so the only contribution
to leading $\hs/\th$ to eq.(\ref{virc}) comes from the octet in
$(\un{8}\otimes\un{8})_A$. Computing then the contraction of the structure
constants with $P^{aa'}_{bb'}(\un{8}_A)$, eq.(\ref{virc}) becomes,
\beq
iM^{aa'bb'}_{\mu_a\mu_{a'}\mu_b\mu_{b'}} \simeq -i 8\pi\a_s\,
{\hs\over\th}\, g_{\mu_a\mu_{a'}} g_{\mu_b\mu_{b'}}\,
f^{ada'} f^{bdb'}\, \ln{\hs\over -\th}\, \a(\th)\, ,\label{vird}
\eeq
which yields the ${\cal O}(\a_s)$ virtual correction to eq.(\ref{tchan}).

We note that between the subleading one-loop corrections we have
neglected there are the self-energy and vertex-correction diagrams
which determine the running of the coupling constant.
Accordingly, in a leading logarithmic treatment of the radiative corrections
$\a_s$ must be regarded as fixed.

\subsection{Multiparton production in multiregge kinematics}
\label{sec:tretwob}

\begin{figure}[hbt]
\vspace*{-4.5cm}
\hspace*{-1.0cm}
\epsfxsize=15cm \epsfbox{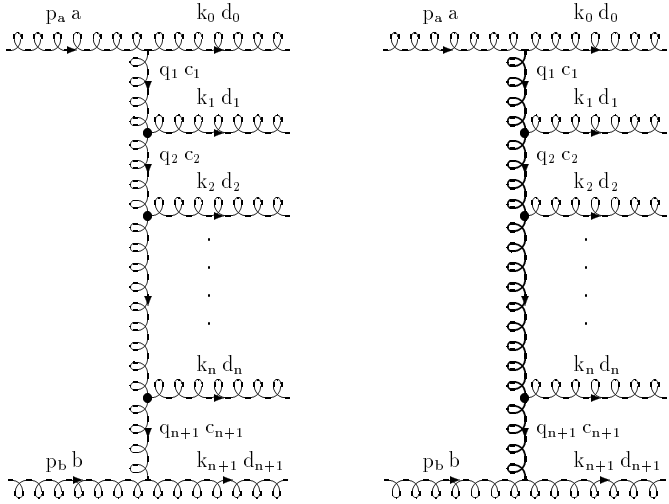}
\vspace*{-10.0cm}
\caption{Multigluon amplitude in multiregge kinematics $a)$ at
tree level and $b)$ with the virtual radiative corrections,
represented by the thicker gluon line. The blobs remind that
non-local effective Lipatov vertices are used for the
gluon emissions along the ladder.}
\label{fig:four}
\end{figure}

In this section, we outline how the results of sect.\ref{sec:tretwo} and
\ref{sec:trenew} are extended to higher orders.
In multiregge kinematics, eq.(\ref{treg}) generalizes to the
emission of an arbitrary number of gluons \cite{BFKL}, namely the tree-level
multigluon amplitude preserves the ladder structure
(Fig.~\ref{fig:four}a), and for the production of $n+2$ gluons it is given by
\bea
iM^{abd_0...d_{n+1}}_{\mu_a\mu_b\mu_0...\mu_{n+1}} &\simeq& 2i \hs
\left(i g_s\, f^{ad_0c_1}\, g_{\mu_a\,\mu_0} \right)\, {1\over \th_1} \nn\\
&\cdot& \left(i g_s\, f^{c_1d_1c_2}\, C^{\mu_1}(q_1,q_2)\right)\,
{1\over \th_2} \nn\\ &\cdot& \label{ntree} \\ &\cdot& \nn\\ &\cdot&
\left(i g_s\, f^{c_nd_nc_{n+1}}\, C^{\mu_n}(q_n,q_{n+1})\right)\,
{1\over \th_{n+1}} \nn\\
&\cdot& \left(i g_s\, f^{bd_{n+1}c_{n+1}}\, g_{\mu_b\,\mu_{n+1}} \right),\nn
\eea
where the Lipatov vertex (\ref{lipver}) is
\beq
C^{\mu}(q_i,q_{i+1}) \simeq \left[(q_i+q_{i+1})^{\mu}_{\perp}\, -\,
\left({\hs_{ai}\over\hs}\,+\,2{\th_{i+1}\over\hs_{bi}}\right) p_b^{\mu}\,
+ \left({\hs_{bi}\over\hs}\,+\,2{\th_i\over\hs_{ai}}\right) p_a^{\mu}\right].
\label{nlipver}
\eeq
A proof of the validity of eq.(\ref{ntree}) may be found in sect.~4 of
ref.\cite{lipat}.

Following ref.~\cite{lip}, we make the ansatz that the leading logarithmic
approximation (LLA) in $\ln(\hs/\th)$ to the virtual radiative corrections,
to all orders
in $\a_s$, is obtained by replacing the propagator for the $i^{th}$ gluon with
\beq
{1\over\th_i} \ra {1\over\th_i}
\left(-{\hs_{i-1,i}\over\th_i}\right)^{\alpha(\th_i)} = {1\over\th_i}
e^{\alpha(\th_i)(y_{i-1}-y_i)}, \label{sud}
\eeq
with $\alpha(\th_i)$ as in eq.(\ref{allv}).
Inserting eq.(\ref{sud}) into eq.(\ref{tchan}) we obtain the
amplitude for the subprocess $g g \ra g g$, with the leading contribution of
the virtual corrections to all orders in $\a_s$,
\beq
iM^{aba'b'}_{\mu_a\mu_b\mu_{a'}\mu_{b'}}\, =\,
-8\pi i\a_s\, f^{aa'c}\, g_{\mu_a\,\mu_{a'}}\, {\hat s\over \hat t}\,
e^{\a(\th)\,(y_a-y_b)}\, f^{bb'c}\, g_{\mu_b\,\mu_{b'}}. \label{vtchan}
\eeq
In eq.(\ref{vird}) we have reproduced the calculation of the
${\cal O}(\a_s^2)$ term of
eq.(\ref{vtchan}) , the ${\cal O}(\a_s^3)$ term has been
computed in ref.\cite{lip}, and in ref.\cite{BFKL} eq.(\ref{vtchan}) has been
conjectured to be valid to all orders in $\a_s$. This entails
that the leading contribution of the virtual radiative corrections to any
order in $\a_s$ has the color-octet structure of one-gluon exchange. Thus,
in the decomposition (\ref{elasgen}) of eq.(\ref{vtchan}) the only scalar
amplitude that contributes to leading $\hs/\th$ is,
\beq
A^{oct}(\hs, \th)\, =\, -8\pi\a_s\, N_c\, {\hat s\over \hat t}\,
e^{\a(\th)\,(y_a-y_b)}. \label{sctchan}
\eeq
Replacing eq.(\ref{sudak}) into eq.(\ref{sud}) we note that the argument
of the exponential factor on the right-hand side is a negative product of a
collinear logarithm and a logarithm of type $\ln(\hs/\th)$. This is an
example of Sudakov form factor. Accordingly the
amplitude (\ref{vtchan}) and the related production rate are infrared
sensitive and vanish as we take $\mu \ra 0$.
This infrared behavior is general in gauge theories and was
first noticed in QED \cite{bd}, where the
probability for having a scattering process without the emission, or with
the emission of a finite number, of soft photons vanishes. It is only after
the inclusion of the real radiative corrections
that a cross section is well defined and finite order by order in
perturbation theory \cite{yfs}.

Inserting then eq.(\ref{sud}) into eq.(\ref{ntree}), we obtain the multigluon
amplitude for the production of $n+2$ gluons, with the leading contribution
of the virtual radiative corrections to all orders in $\a_s$
\bea
iM^{abd_0...d_{n+1}}_{\mu_a\mu_b\mu_0...\mu_{n+1}} &\simeq& 2i \hs
\left(i g_s\, f^{ad_0c_1}\, g_{\mu_a\,\mu_0} \right)\, {1\over \th_1}
e^{\a(\th_1)\,(y_0-y_1)} \nn\\
&\cdot& \left(i g_s\, f^{c_1d_1c_2}\, C^{\mu_1}(q_1,q_2)\right)\,
{1\over \th_2} e^{\a(\th_2)\,(y_1-y_2)} \nn\\ &\cdot& \label{ngluon} \\
&\cdot& \nn\\ &\cdot&
\left(i g_s\, f^{c_nd_nc_{n+1}}\, C^{\mu_n}(q_n,q_{n+1})\right)\,
{1\over \th_{n+1}} e^{\a(\th_{n+1})\,(y_n-y_{n+1})} \nn\\
&\cdot& \left(i g_s\, f^{bd_{n+1}c_{n+1}}\, g_{\mu_b\,\mu_{n+1}} \right).\nn
\eea
Eq.(\ref{ngluon}) has been verified for three-gluon production at one loop
in ref.~\cite{BFKL}. A proof of its validity to all orders in $\a_s$ may be
found in the appendix of
ref.~\cite{lipat}. We will follow, though, the original approach of
ref.~\cite{BFKL}, namely we take eq.(\ref{ngluon}) as a corollary of the
ansatz (\ref{sud}) and we prove its self-consistency by using eq.(\ref{ngluon})
to derive the elastic
amplitude with the virtual radiative corrections in LLA to all orders in
$\a_s$, and by showing that it coincides with eq.(\ref{vtchan}).

\subsection{Partial wawe amplitudes}
\label{sec:tretre}

In this section we derive dispersion relations
which will let us reconstruct the elastic
amplitude with all the virtual radiative corrections (\ref{vtchan}).


First, we decompose the scalar amplitude $A^T(\hs, \th)$ introduced in
eq.(\ref{elasgen})in $\th$-channel partial wawe amplitudes
\beq
A^T(\hs, \th) = \sum_l\, (2l+1)\, A^T_l(\hs, \th)\, P_l(z_t), \label{pwawe}
\eeq
where $l$ is the angular momentum, $P_l(z)$ are Legendre polynomials
and $z_t = -\cos\theta_t$ is the scattering angle in the $\th$-channel
physical region, whose Mandelstam invariants are obtained from eq.(\ref{scat})
by crossing the channels $\hs$ and $\th$,
\bea
\hat s &=& -{\hat t \over 2} (1-\cos\theta_t), \label{tscat} \\
\hat u &=& -{\hat t \over 2} (1+\cos\theta_t). \nn
\eea
Using the invariance of amplitude (\ref{elasgen}) under the crossing
symmetry $\hs \leftrightarrow \hat u$,
\beq
M^{aba'b'}_{\mu_a\mu_b\mu_{a'}\mu_{b'}}(\hs, \th, \hat u)\, =\,
M^{ab'a'b}_{\mu_a\mu_{b'}\mu_{a'}\mu_b}(\hat u, \th, \hs)\, ,
\eeq
and the parity of the projectors (\ref{parit}),
we obtain the parity of the scalar amplitudes
under $\hs \leftrightarrow \hat u$ crossing
\beq
A^T(-z_t,\th) = (-1)^T\, A^T(z_t,\th)\, , \label{par}
\eeq
with the change of sign of the scattering angle, $z_t \leftrightarrow -z_t$
under $\hs \leftrightarrow \hat u$ crossing, as seen from eq.(\ref{tscat}).
Next, we write the elastic amplitude (\ref{elasgen}) through a dispersion
relation, i.e. as an integral over the amplitude singularities,
which are branch cuts over the real axis of the $\hs$ complex plane,
$-\th \le \hs < \infty$ for the physical $\hs$ channel, and
$-\th \le \hat u < \infty$ for the physical $\hat u$ channel \cite{bj},
\beq
A(\hs, \th) = \int_{-\infty}^0 {ds'\over 2\pi i}\, {{\rm Disc}\, A(s',\th)
\over s'-\hs}\, +\, \int_{-\th}^{\infty} {ds'\over 2\pi i}\, {{\rm Disc}\,
A(s',\th)\over s'-\hs}\, , \label{disp}
\eeq
with
\beq
{\rm Disc}\, A(s',\th) = A(s'+i\epsilon, \th) - A(s'-i\epsilon, \th)\, ,
\label{disc}
\eeq
and where we have used $\hat u = -\hs -\th$, and $\th < 0$. Since, from
eq.(\ref{tscat}),
\beq
z_t = -\left(1 + {2\hs\over\th}\right), \label{zt}
\eeq
the dispersion relation (\ref{disp}) may be written as an integral over the
complex plane of $z_t$,
\beq
A(\hs, \th) = \int_{-\infty}^{-1} {dz_t'\over 2\pi i}\, {{\rm Disc}\,
A(z_t',\th) \over z_t'-z_t}\, +\, \int_1^{\infty} {dz_t'\over 2\pi i}\,
{{\rm Disc}\, A(z_t',\th)\over z_t'-z_t}\, . \label{dispz}
\eeq
We note that in the $\th$-channel physical region (\ref{tscat}), where $z_t$
is the scattering angle, $-1 \le z_t \le 1$, while the singularities are
at $z_t < -1$ and $z_t > 1$, where the
$\th$ channel is unphysical (\ref{dispz}). So at $-1 \le z_t \le 1$
we may invert the partial-wawe expansion (\ref{pwawe}), to obtain the
amplitude for the $l^{th}$ wawe
\beq
A^T_l(\hs, \th)\, =\, {1\over 2} \int_{-1}^1 dz_t\, P_l(z_t)\, A^T(\hs, \th)\,.
\label{invp}
\eeq
Introducing the Legendre function, associated to the Legendre polynomial
\beq
Q_l(z')\, =\, {1\over 2} \int_{-1}^1 {dz\over z'-z} P_l(z)\, ,
\eeq
replacing the dispersion relation (\ref{dispz}) into eq.(\ref{invp}),
and using the parity under $\hs \leftrightarrow \hat u$ crossing,
\bea
Q_l(-z_t) &=& (-1)^{l+1}\, Q_l(z_t)\, , \\
{\rm Disc}\, A^T(-z_t,\th) &=& (-1)^{T+1}\, {\rm Disc}\, A^T(z_t,\th)\, ,
\nn
\eea
the amplitude for the $l^{th}$ wawe becomes
\beq
A^T_l(\hs, \th)\, =\, \left[1 + (-1)^{l+T}\right]\, \int_1^{\infty}
{dz'\over 2\pi i}\, Q_l(z')\, {\rm Disc}\,A^T(z',\th)\, . \label{invpb}
\eeq
Next, we introduce the Sommerfeld-Watson representation of the
amplitude $A^T(\hs, \th)$ in the complex plane of the angular momentum $l$
\cite{pdb},
\beq
A^T(\hs, \th)\, =\, \int_{\delta-i\infty}^{\delta+i\infty} {dl\over 2i}\,
(2l+1)\, A^T_l(\hs, \th)\, {P_l(-z_t)\over \sin{\pi l}}, \label{sw}
\eeq
where the integral is done along a path parallel to the imaginary axis,
to the right of all the singularities. All that we have considered so far in
this section describes general properties of the analiticity of the scattering
amplitude. Now, we consider the high-energy limit $\hs\ra\infty$ at fixed
$\th$. From eq.(\ref{zt}) we have,
\beq
z_t \ra -{2\hs\over\th}\, ,
\eeq
and we take the asymptotic value for large
$z_t$ of the Legendre polynomials and their associated functions \cite{abr}
\bea
P_l(z) &\ra& {1\over\sqrt{\pi}}\, {\Gamma\left(l+{1\over 2}\right)\over
\Gamma(l+1)}\, (2z)^l\, , \label{asym} \\
Q_l(z) &\ra& \sqrt{\pi}\, {\Gamma(l+1)\over\Gamma\left(l+{3\over 2}\right)}\,
(2z)^{-(l+1)}\, . \nn
\eea
Replacing the amplitude for the $l^{th}$ wawe (\ref{invpb}) into the
Sommerfeld-Watson representation of the amplitude (\ref{sw}), and using
the asymptotics (\ref{asym}), we obtain
\beq
A^T(\hs, \th)\, =\, -{1\over 4\pi} \int_{\delta-i\infty}^{\delta+i\infty} dl\,
{(-1)^l + (-1)^T\over \sin{\pi l}} e^{ly} {\cal F}^T_l(\th)\, , \label{swamp}
\eeq
where ${\cal F}^T_l(\th)$ is the Laplace transform of the discontinuity of the
amplitude
\beq
{\cal F}^T_l(\th)\, =\, \int_0^{\infty} dy\, e^{-ly}\, {\rm Disc}\,
A^T(z_t,\th)\, ,
\label{lapl}
\eeq
with $y = \ln{(z_t/2)}$.

\subsection{The BFKL equation}
\label{sec:trefour}

Our goal is now to use the $\hs$-channel unitarity and
the multigluon amplitude with exchange of one gluon
in the $\th$ channel (\ref{ngluon}) to evaluate the discontinuity of the
amplitude for exchange of two gluons in the $\th$ channel in a physical gauge.
{}From that by using the dispersion relations computed above the elastic
amplitude with all the virtual radiative corrections (\ref{vtchan}) will be
derived.

\begin{figure}[hbt]
\vspace*{-4.5cm}
\hspace*{-3.0cm}
\epsfxsize=15cm \epsfbox{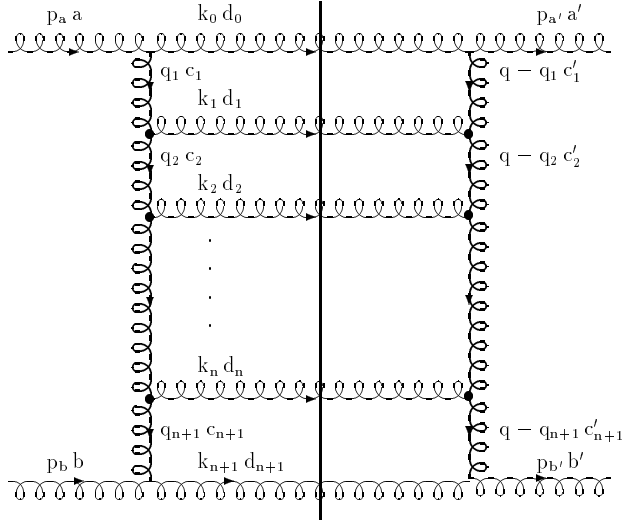}
\vspace*{-10.0cm}
\caption{${\cal O}(\a_s^{n+2})$ contribution to the discontinuity of the
elastic amplitude with two-gluon ladder exchange in the $\th$ channel.}
\label{fig:five}
\end{figure}

In order to evaluate the discontinuity of the amplitude, we
compute and sum over $n$ the cut diagram of Fig.\ref{fig:five}.
In evaluating the diagram at $n+1$ loops, we must put the internal
legs $k_i$ on mass shell, by replacing the $n+2$ internal propagators with the
cut ones \cite{ram}
\beq
{i\over k_i^2} \ra 2\pi\, \delta(k_i^2)\, .\label{onsh}
\eeq
Multiplying then the mass-shell factors (\ref{onsh}) by the $n+1$ loop-momentum
integrals, we obtain the phase space for the production of $n+2$ partons
\beq
{\cal P}_{n+2} \equiv \prod_{i=0}^n \int {d^4 k_i\over (2\pi)^4} \,
\prod_{j=0}^{n+1} 2\pi\,
\delta(k^2_j) \, = \, \prod_{i=0}^{n+1} \int {dy_i\, d^2 k_{i\perp}\over 4\pi
(2\pi)^2}\, (2\pi)^4 \, \delta^4(p_a + p_b - \sum_{i=0}^{n+1} k_i) \label{phsp}
\eeq
where we have inserted the identity $\int d^4 k_{n+1} \delta^4(p_a + p_b -
\sum_{i=0}^{n+1} k_i) = 1$, and used eq.(\ref{ph}).
Now we need to consider the phase space
in the multiregge kinematics. This has already been computed for three-parton
production in eq.(\ref{mrtps}). Since light-cone momentum conservation in the
multiregge kinematics (\ref{ykin}) has the same form for three or more
partons, eq.(\ref{mrtps}) trivially generalizes to $n+2$ partons, and the
phase space (\ref{phsp}) becomes,
\beq
{\cal P}_{n+2} = \int {1\over 2\hs}\, {d^2 k_{0\perp}\over (2\pi)^2}\,
\left(\prod_{i=1}^n {dy_i\, d^2 k_{i\perp}\over 4\pi (2\pi)^2}\right)\,
{d^2 k_{n+1\perp}\over (2\pi)^2}\, (2\pi)^2 \, \delta^2\left(\sum_{i=0}^{n+1}
k_{i\perp}^2\right)\, . \label{mrnps}
\eeq
Then using eq.(\ref{ngluon}) the discontinuity of the amplitude of
Fig.\ref{fig:five} is
\bea
& & {\rm Disc} \left[iM^{aba'b'}_{\mu_a\mu_b\mu_{a'}\mu_{b'}}(\hs,\th)\right]\,
=\nn\\ & & \sum_{n=0}^{\infty} \int {1\over 2\hs}\, {d^2 k_{0\perp}\over
(2\pi)^2}\,\left(\prod_{i=1}^n {dy_i\, d^2 k_{i\perp}\over 4\pi (2\pi)^2}
\right)\, {d^2 k_{n+1\perp}\over (2\pi)^2}\, (2\pi)^2 \, \delta^2\left(
\sum_{i=0}^{n+1} k_{i\perp}^2\right) \nn\\ &\cdot& (2i \hs)^2\,
\delta_{\perp}^{\mu_a\mu_{a'}}\, \delta_{\perp}^{\mu_b\,\mu_{b'}}\,
\left(i g_s\, f^{ad_0c_1}\right)\, \left(i g_s\, f^{c'_1d_0a'}\right) \nn\\
&\cdot& {1\over \th_1} e^{\a(\th_1)\,(y_0-y_1)}\, {1\over \th'_1}
e^{\a(\th'_1)\,(y_0-y_1)} \nn\\ &\cdot& \left(i g_s\, f^{c_1d_1c_2}\right)\,
\left(i g_s\, f^{c'_1d_1c'_2}\right)\, C^{\mu_1}(q_1,q_2) (-g_{\mu_1\mu'_1})
C^{\mu'_1}(q-q_1,q-q_2) \nn\\ &\cdot&
{1\over \th_2} e^{\a(\th_2)\,(y_1-y_2)}\, {1\over \th'_2}
e^{\a(\th'_2)\,(y_1-y_2)} \nn\\ &\cdot& \label{ndisc} \\
&\cdot& \nn\\ &\cdot&
\left(i g_s\, f^{c_nd_nc_{n+1}}\right)\, \left(i g_s\, f^{c'_nd_nc'_{n+1}}
\right)\, C^{\mu_n}(q_n,q_{n+1}) (-g_{\mu_n\mu'_n})
C^{\mu'_n}(q-q_n,q-q_{n+1}) \nn\\ &\cdot&
{1\over \th_{n+1}} e^{\a(\th_{n+1})\,(y_n-y_{n+1})}\,
{1\over \th'_{n+1}} e^{\a(\th'_{n+1})\,(y_n-y_{n+1})} \nn\\
&\cdot& \left(i g_s\, f^{bd_{n+1}c_{n+1}}\right)
\left(i g_s\, f^{c'_{n+1}d_{n+1}b'}\right)\, ,\nn
\eea
where $\th = q^2$ is the momentum transfer in the elastic scattering and
$\th'_i = (q-q_i)^2$. In eq.(\ref{ndisc})
we have used the phase space (\ref{mrnps}), and we have done the
sum over the helicities of the gluons emitted from the helicity-conserving
vertices using eq.(\ref{tran}), and the ones over the $n$ intermediate gluons
using eq.(\ref{feyn}) since the Lipatov vertices are gauge invariant
(\ref{gauge}). The Lipatov vertex $C^{\mu'_i}(q-q_i,q-q_{i+1})$ in
eq.(\ref{ndisc}) may be obtained either by direct construction as in
sect.~\ref{sec:tretwo} or by simply inverting the sign of the momentum $k_i$ in
eq.(\ref{nlipver}),
\beq
C^{\mu}(q-q_i,q-q_{i+1}) \simeq \left[\left((q-q_i)\,+\,(q-q_{i+1})\right)^
{\mu}_{\perp}\, +\,
\left({\hs_{ai}\over\hs}\,+\,2{\th'_{i+1}\over\hs_{bi}}\right) p_b^{\mu}\,
- \left({\hs_{bi}\over\hs}\,+\,2{\th'_i\over\hs_{ai}}\right) p_a^{\mu}\right].
\label{nlipverc}
\eeq
Then using eq.(\ref{tman}), (\ref{nlipver}), (\ref{nlipverc}) and the kinematic
constraint $\hs_{ai} \hs_{bi} = k_{i\perp}^2 \hs$ from eq.(\ref{invb}), with
$k_{i\perp}^2 = (q_i - q_{i+1})_{\perp}^2$,
the contraction of the Lipatov vertices becomes
\bea
C^{\mu}(q_i,q_{i+1})\, C_{\mu}(q-q_i,q-q_{i+1})\, &=& -2 \left[q_{\perp}^2\,
-\, {(q-q_i)_{\perp}^2 q_{i+1\perp}^2 + (q-q_{i+1})_{\perp}^2 q_{i\perp}^2
\over (q_i - q_{i+1})_{\perp}^2}\right] \nn\\
&\equiv& -2 {\cal K}(q_i,q_{i+1})\, . \label{lipc}
\eea
Next, using eq.(\ref{elasgen}) we consider the scalar part
${\rm Disc} A^T(\hs, \th)$ of the discontinuity
(\ref{ndisc}). The contraction of the
color projectors (\ref{octsin}) with the structure
constants in eq.(\ref{ndisc}) yields the color factor $C_T^{n+2}$, with
\beq
C_T = \left\{ \begin{array}{ll} N_c & \mbox{for the {\sl singlet}},\\
{N_c/2} & \mbox{for the {\sl octet}}, \end{array}
\right. \label{cf}
\eeq
where we have used the Jacobi identity (\ref{jacob}) in the contraction with
the octet projector. So the scalar part of the discontinuity
(\ref{ndisc}) may be written using eq.(\ref{lipc}) and (\ref{cf}) as,
\bea
{\rm Disc} A^T(\hs, \th) &=& \sum_{n=0}^{\infty} \left(-g_s^2 C_T
\right)^{n+2} \int \prod_{i=1}^n {dy_i\over 4\pi}\,
\prod_{j=1}^{n+1} {d^2q_{j\perp}\over (2\pi)^2}\, \nn\\ &\cdot& 2i\, \hs\,
\prod_{l=1}^{n+1} {1\over \th_l \th'_l}\, e^{(y_{l-1}-y_l)
[\a(\th_l)+\a(\th'_l)]}\, \prod_{m=1}^n 2{\cal K}(q_m,q_{m+1})\, ,\label{ndisd}
\eea
where we have used the conservation of the transverse momentum and we have
changed integration variables from the transverse components of the produced
gluons to the ones of the gluons exchanged in the $\th$ channel. In
eq.(\ref{ndisd}) we have $n$ integrations over the rapidities of the gluons
emitted within the ladder, while in the integrand there are $n+1$ rapidity
differences. To disentangle the integrations, we take the Laplace transform
(\ref{lapl}) of eq.(\ref{ndisd}) with respect to the rapidity difference
$y = y_0 - y_{n+1} = \ln{(-\hs/\th)}$, and change the integrations
over the rapidities $y$ and $y_i$, with $i=1,...,n$ to the ones over the
rapidity differences $y_{i-1} - y_i$, with $i=1,...,n+1$. Then we perform the
integrations, with boundaries given by the strong rapidity ordering
(\ref{mreg}), and we obtain
\bea
{\cal F}^T_l(\th) &=& -2i\, \th\, (4\pi\a_s)^2\, C_T^2\, \sum_{n=0}^{\infty}
\int \prod_{j=1}^{n+1} {d^2q_{j\perp}\over (2\pi)^2}\, \nn\\
&\cdot& {1\over \th_1 \th'_1}\, {1\over l - 1- \a(\th_1) - \a(\th'_1)} \nn\\
&\cdot& (-2\a_s C_T)\, {\cal K}(q_1,q_2) \nn\\ &\cdot&
{1\over \th_2 \th'_2}\, {1\over l - 1- \a(\th_2) - \a(\th'_2)}
\nn\\ &\cdot& \label{nlapl} \\ &\cdot& \nn\\ &\cdot& (-2\a_s C_T)\,
{\cal K}(q_n,q_{n+1}) \nn\\ &\cdot& {1\over \th_{n+1} \th'_{n+1}}\,
{1\over l - 1- \a(\th_{n+1}) - \a(\th'_{n+1})}\, . \nn
\eea
This may be written as a recursive relation,
\beq
{\cal F}^T_l(\th)\, =\, -2i\, \th\, (4\pi\a_s)^2\, C_T^2\, \int {d^2q_{1\perp}
\over (2\pi)^2}\, {1\over q_{1\perp}^2 (q-q_1)_{\perp}^2}\, f_l^T(q_1,\th)\, ,
\label{recur}
\eeq
where the function $f_l^T(q,\th)$ satisfies the integral equation,
\beq
\left[l - 1- \a(\th_1) - \a(\th'_1)\right] f_l^T(q_1,\th)\, =\, 1 - 2\a_s C_T
\int {d^2q_{2\perp}\over (2\pi)^2}\, {{\cal K}(q_1,q_2)\over q_{2\perp}^2
(q-q_2)_{\perp}^2}\, f_l^T(q_2,\th)\, ,\label{bfkle}
\eeq
with $\th_i=-q_{i\perp}^2$ and $\th'_i=-(q-q_i)_{\perp}^2$.
Eq.(\ref{bfkle}) is the BFKL integral equation \cite{BFKL}, describing
the gluon-ladder evolution in the LLA of $\ln{(\hs/\th)}$. The function
${\cal K}(q_1,q_2)$, defined as the contraction of the Lipatov vertices
(\ref{lipc}), describes the contribution of the real radiative corrections
and forms
the kernel of the BFKL equation. The contribution of the virtual radiative
corrections $\a(\th)$ appears on the left-hand side of
eq.(\ref{bfkle}).

Next, we are going to solve the BFKL equation for the color-octet
exchange. Replacing then eq.(\ref{cf}) and the explicit form (\ref{allv}) of
$\a(\th)$ into eq.(\ref{bfkle}), this reduces to
\beq
(l - 1)\, f_l^{oct}(q_1,\th)\, =\, 1 - \a_s N_c\, q_{\perp}^2 \int {d^2\kt
\over (2\pi)^2}\, {1\over\kt^2 (q-k)_{\perp}^2}\, f_l^{oct}(k,\th)\,
,\label{red}
\eeq
which admits the solution
\beq
f_l^{oct}(q_1,\th)\, =\, f_l^{oct}(k,\th)\, =\, {1\over l - 1 - \a(\th)}\,
.\label{ocs}
\eeq
The solution is unique, since eq.(\ref{bfkle}) is inhomogeneous, and it is
constant with respect to the functional dependence from its first argument.
Replacing eq.(\ref{ocs})
into the Laplace transform (\ref{recur}) and using eq.(\ref{allv}) and
(\ref{cf}), we obtain
\beq
{\cal F}^{oct}_l(\th)\, =\, -8\pi^2 i\, N_c\,\a_s\, {\a(\th)\over l - 1
-\a(\th)}\,
.\label{ocd}
\eeq
Replacing eq.(\ref{ocd}) into the Sommerfeld-Watson representation of
amplitude (\ref{swamp}) and using the parity factor (\ref{cp}), the scalar
part of the amplitude for octet exchange becomes
\beq
A^{oct}(\hs,\th)\, =\, 4\pi\, N_c\,\a_s\, {\pi\a(\th)\over \sin{\pi\a(\th)}}\,
\left(1 + e^{i\pi\a(\th)}\right)\, \left(\hs\over -\th\right)^{1+\a(\th)}\,
,\label{ocamp}
\eeq
where the integral over the complex plane of $l$ has yielded a Regge pole at
$l = 1 +\a(\th)$, which then gives the Regge trajectory $\hs^{1+\a(\th)}$.
Since $\a(\th=0) = 0$, the intercept is at $l = 1$, which corresponds to a
reggeized gluon \cite{BFKL}, \cite{pdb}.

Even though we started with a discontinuity
(\ref{ndisc}) at ${\cal O}(\a_s^2)$, that computes the radiative
corrections due the insertion of a two-gluon ladder, the octet solution
(\ref{ocamp}) is ${\cal O}(\a_s)$, because the octet
appears already at the one-gluon exchange level.

Finally, we note that in the multiregge region (\ref{mreg}), where the
collinear
logarithms are not supposed to be large, $\a_s \ln{(q_{\perp}^2/\mu^2)} \ll 1$.
Then from eq.(\ref{sudak}) we obtain that $\pi |\a(\th)| \ll 1$.
Taking this limit in eq.(\ref{ocamp}), we find it in agreement with
eq.(\ref{sctchan}), proving thereby the self-consistency of the ansatz
(\ref{sud}) for eq.(\ref{vtchan}) \cite{BFKL}.

\subsection{The pomeron solution}
\label{sec:trefive}

We are now interested to study the singlet solution of the BFKL equation
(\ref{bfkle}), since it can be related via the optical theorem to the total
cross section for one-gluon exchange in the $\th$ channel.
Let us write first the function $f_l^{sing}(q_1,\th)$
in differential form,
\beq
f_l^{sing}(q_1,\th)\, =\, \int {d^2\kt\over (2\pi)^2} {\bar f}_l(q_1,k,\th)\,
.\label{baf}
\eeq
Replacing it into eq.(\ref{bfkle}) and using eq.(\ref{cf}), we obtain the
BFKL equation for color-singlet exchange,
\bea
& & \left[l - 1- \a(\th_1) - \a(\th'_1)\right] {\bar f}_l(q_1,k,\th)\, =\nn\\
& & (2\pi)^2\,\delta^2(q_1-k)\, -\, 2\a_s N_c\,
\int {d^2q_{2\perp}\over (2\pi)^2}\, {{\cal K}(q_1,q_2)\over q_{2\perp}^2
(q-q_2)_{\perp}^2}\, {\bar f}_l(q_2,k,\th)\, .\label{bfkls}
\eea
Since by pomeron exchange it is meant conventionally the exchange of
something with quantum numbers of the vacuum, the exchange of a
two-gluon ladder in a color-singlet configuration defines the exchange of
a perturbative QCD pomeron.

The kernel ${\cal K}(q_1,q_2)$ (\ref{lipc}) is regular as $q^2_{2\perp}
\ra\infty$, making the right-hand side of eq.(\ref{bfkls}) ultraviolet finite.
There are ultraviolet divergences as $q^2_{1\perp}\ra\infty$ in the virtual
radiative corrections (\ref{sudak}) in the left-hand side of eq.(\ref{bfkls}),
but they cancel in the singlet solution (\ref{recur}).
As for the infrared behavior of eq.(\ref{bfkls}), the kernel vanishes at
$q_{2\perp}^2 = 0$, or at $(q-q_2)_{\perp}^2 = 0$, where accordingly the
right-hand side of eq.(\ref{bfkls}) is regular. The kernel is singular at
$q_1^2 = q_2^2$, however making the substitution
\beq
\int d^2q_2 {q_{1\perp}^2\over q_{2\perp}^2 (q_1-q_2)_{\perp}^2}\, =\,
2\, \int d^2q_2 {q_{1\perp}^2\over \left[q_{2\perp}^2 + (q_1-q_2)_{\perp}^2
\right]\, (q_1-q_2)_{\perp}^2}\, ,\label{iden}
\eeq
in the virtual radiative-corrections terms of eq.(\ref{bfkls}), it is easy
to see that the singularity at $q_1^2 = q_2^2$ cancels between the virtual
and real radiative corrections. Still, eq.(\ref{bfkls}) has
infrared divergences as $q_1^2\ra 0$ in the virtual
radiative corrections. Besides, at $q_{1\perp}^2 = 0$ or
$(q-q_1)_{\perp}^2 = 0$ the kernel vanishes, and after using
eq.(\ref{baf}), eq.(\ref{bfkls}) reduces to eq.(\ref{red}), thus
the two-gluon ladder reduces to a one-gluon ladder, which is infrared
sensitive as we know from sect.3.4 and 3.6.
However, it is possible to show that in particular
cases, like for the exchange
of a singlet gluon ladder between colorless objects, the solution of
eq.(\ref{bfkls}) has no infrared divergences at all \cite{conf}.

Since we are interested in the forward scattering amplitude and in the total
cross section, we look for the solution of eq.(\ref{bfkls}) at $q = 0$.
{}From eq.(\ref{csq}) and (\ref{lipc}) we obtain
\beq
{\cal K}(q_1,q_2)|_{q=0}\, =\, -2\, {q_{1\perp}^2 q_{2\perp}^2
\over (q_1-q_2)_{\perp}^2}\, .\label{zk}
\eeq
Doing then the substitution of variables,
\beq
f_l(q_1,k)\, =\, {1\over 8\pi^2}\, {\kt^2\over q_{1\perp}^2}\,
{\bar f}_l(q_1,k,\th=0)\, ,\label{subst}
\eeq
the BFKL equation for singlet exchange at $\th = 0$ becomes
\bea
& & (l-1)\, f_l(q_1,k)\, =\label{bfkla}\\ & & {1\over 2}\,\delta^2(q_1-k)\,
+\, {\a_s N_c\over\pi^2}
\int d^2q_{2\perp}\, {1\over (q_1-q_2)_{\perp}^2}\,\left[f_l(q_2,k) -
{q_{1\perp}^2\over 2q_{2\perp}^2}\, f_l(q_1,k)\right]\, .\nn
\eea
This is a inhomogeneous integral equation with a self-adjoint kernel, given by
the real radiative corrections to the discontinuity of the amplitude.
As we have done for eq.(\ref{bfkls}),
in order to regulate the divergence at $q_1^2 = q_2^2$ we
make the substitution (\ref{iden}) in the virtual radiative-corrections term
on the right-hand side of eq.(\ref{bfkla}) (cf. Appendix A),
then the BFKL equation at $\th = 0$ reads
\bea
& & (l-1)\, f_l(q_1,k)\, =\label{bfklb}\\ & & {1\over 2}\,\delta^2(q_1-k)\,
+\, {\a_s N_c\over \pi^2} \int d^2q_{2\perp}\, {1\over (q_1-q_2)_{\perp}^2}\,
\left[f_l(q_2,k) - {q_{1\perp}^2\over q_{2\perp}^2 + (q_1 - q_2){\perp}^2}\,
f_l(q_1,k)\right]\, .\nn
\eea
The homogeneous equation associated to eq.(\ref{bfklb}), may be written as
\bea
& & (l-1)\, f_l(q_1,k)\, =\label{homog}\\ & &
{\a_s N_c\over \pi^2} \int d^2q_{2\perp}\, \left\{ {f_l(q_2,k)\over
(q_1-q_2)_{\perp}^2}\, -\,{q_{1\perp}^2\over q_{2\perp}^2}
\left[{1\over (q_1 - q_2)_{\perp}^2}\, -\, {1\over q_{2\perp}^2 +
(q_1 - q_2)_{\perp}^2}\right]\, f_l(q_1,k)\right\}\, ,\nn
\eea
and admits a solution as a Fourier series
\beq
f_l(q_1,k)\, =\, \sum_{n=-\infty}^{\infty}\, \int_{-\infty}^{\infty} d\nu\,
a(\nu,n)\, e^{i\nu(\lambda_1-\lambda)}\, e^{in(\phi_1-\phi)}\, ,\label{sol}
\eeq
with
$\lambda_1 = \ln(q_1^2/m^2)$, $\lambda = \ln(k^2/m^2)$, $m^2$ a scale
factor, and $\phi_1-\phi$ the
azimuthal angle between the vectors $\kt$ and $q_{1\perp}$.
Using then the integral representation of the $\delta$-function, the
inhomogeneous term in eq.(\ref{bfklb}) can be expanded as
\beq
\delta^2(q_1-k)\, =\, {1\over (\kt^2 q_{1\perp}^2)^{1/2}}\, {1\over 2\pi^2}\,
\sum_{n=-\infty}^{\infty} \int_{-\infty}^{\infty} d\nu\,
e^{i\nu(\lambda_1-\lambda)}\, e^{in(\phi_1-\phi)}\, .\label{nhom}
\eeq
Replacing eq.(\ref{nhom}) into eq.(\ref{bfklb}) and calling $\omega(\nu,n)$
the eigenvalue of the homogeneous equation (\ref{homog}),
we obtain the condition
\beq
(l-1)\, a(\nu,n)\, =\, {1\over (\kt^2 q_{1\perp}^2)^{1/2}}\, {1\over
(2\pi)^2}\, +\, \omega(\nu,n) a(\nu,n)\, ,
\eeq
which fixes the coefficient $a(\nu,n)$ in the solution (\ref{sol}),
\beq
f_l(q_1,k)\, =\, {1\over (2\pi)^2}\, {1\over (\kt^2 q_{1\perp}^2)^{1/2}}\,
\sum_{n=-\infty}^{\infty} \int_{-\infty}^{\infty} d\nu\,
{1\over l -1 -\omega(\nu,n)}\,
e^{i\nu(\lambda_1-\lambda)}\, e^{in(\phi_1-\phi)}\, .\label{solb}
\eeq
To find the spectrum of eigenvalues, we replace the solution (\ref{solb}) into
the homogeneous equation (\ref{homog}), and obtain
\bea
\omega(\nu,n)\, &=& {\a_s N_c\over\pi}\, \left[2\,{\rm Re}\int_0^1 dx\,
{x^{{|n|-1\over 2}+i\nu} \over 1-x} - 2 \int_0^1 dx\, {1\over 1-x} \right.
\nn\\ & & \left. -\,\int_0^1 dx\, {1\over x}\, +\, \int_0^1 dx\, {1\over x
\sqrt{1+4x^2}}\, +\, \int_0^1 dx\, {1\over \sqrt{x^2+4}}\right]\, ,\label{spec}
\eea
with
\beq
x =\, \left\{ \begin{array}{ll}
q_2^2/q_1^2 & \mbox{for $q_2^2 < q_1^2$}\, ,\\
q_1^2/q_2^2 & \mbox{for $q_2^2 > q_1^2$}\, .\end{array}
\right. \label{xvar}
\eeq
The last three terms in eq.(\ref{spec}) cancel out,
and introducing the logarithmic derivative of the $\Gamma$ function
\beq
{d\ln{\Gamma(y)}\over dy}\, =\, \psi(y)\, =\, \int_0^1 dx {x^{y-1}-1\over x-1}
- \gamma\, ,\label{gam}
\eeq
with $\gamma = -\psi(1) = 0.577215...$ the Euler-Mascheroni constant, the
eigenvalue (\ref{spec}) becomes
\beq
\omega(\nu,n)\, =\, -2{\a_s N_c\over\pi}\, {\rm Re}\left[\psi\left({|n|+1
\over 2} +i\nu\right) -\psi(1)\right]\, .\label{om}
\eeq
$\omega(\nu,n)$ is regular at both the integration limits (\ref{gam}),
entailing that eq.(\ref{bfklb}) is regular at
$q_2^2 = 0$, $q_2^2 = q_1^2$, and $q_2^2\ra\infty$, i.e. that
all the infrared and ultraviolet divergences cancel out.

Next, we use the solution (\ref{solb}) in the Laplace transform of the
discontinuity of
the amplitude (\ref{recur}). The right-hand side of eq.(\ref{recur}) is
proportional to $\th$, so it seems to vanish. However, the factor $\th$ was
due to replacing $\hs = -\th e^{y_0-y_{n+1}}$ into eq.(\ref{ndisd}), and for
the $\th=0$ case we should have rather done the substitution
$\hs = \kt^2 e^{y_0-y_{n+1}}$, with $\kt^2$ as in eq.(\ref{mreg}). Doing so,
and using eq.(\ref{cf}), (\ref{baf}) and (\ref{subst}), the Laplace transform
of the discontinuity for singlet exchange becomes
\beq
{\cal F}^{sing}_l(\th=0)\, =\, 16i\,\kt^2\,\a_s^2\, N_c^2\,\int d^2\kta\,
d^2\ktb\, {1\over \kta^2\ktb^2} f_l(k_a,k_b)\, ,\label{sindis}
\eeq
with $f_l(k_a,k_b)$ as in eq.(\ref{solb}), with eigenvalue (\ref{om}). Before
taking the inverse Laplace transform of eq.(\ref{sindis}),
\beq
{\rm Disc} A(\hs,\th)\, =\, \int {dl\over 2\pi i} e^{ly} {\cal F}_l(\th)\, ,
\label{invd}
\eeq
it is convenient to examine the eigenvalue (\ref{om}) to determine where
the leading singularity at large $y$ is.
The leading contribution to $\omega(\nu,n)$ comes
from $n=\nu=0$ (cf. Appendix B), and for small $\nu$ the eigenvalue
admits the expansion (\ref{omex})
\beq
\omega(\nu,n=0)\, =\, 2{\a_s N_c\over\pi}\, \left(2\ln{2} - 7\zeta(3)\,\nu^2\,
+\,...\right)\, .\label{omexp}
\eeq
Replacing it into eq.(\ref{solb}), and evaluating the integral over the
complex plane of $\nu$, we obtain the leading contribution to the solution for
the singlet,
\beq
f_l(k_a,k_b)\, \simeq\, {1\over (2\pi)^2}\, {1\over (\kta^2 \ktb^2)^{1/2}}\,
{\pi\over \left[B(l-1-A)\right]^{1/2}}\, e^{-\nu_0 |\ln({\kta^2/\ktb^2})|}\, .
\label{asyso}
\eeq
with
\beq
A\, =\, 4\ln{2}\, {\a_s N_c\over\pi}\, ;\qquad B\, =\, 14\zeta(3)\, {\a_s N_c
\over\pi}\, ;\qquad \nu_0\, =\, \left({l-1-A\over B}\right)^{1/2}\, .
\label{scoef}
\eeq
Replacing it into the discontinuity (\ref{invd}), we see that in the
complex plane of $l$ the leading singularity of the pomeron solution is a
branch cut, extending from $-\infty$ up to $l=1+A$ \cite{BFKL2}.

\begin{figure}[hbt]
\vspace*{-5.5cm}
\hspace*{-0.5cm}
\epsfxsize=15cm \epsfbox{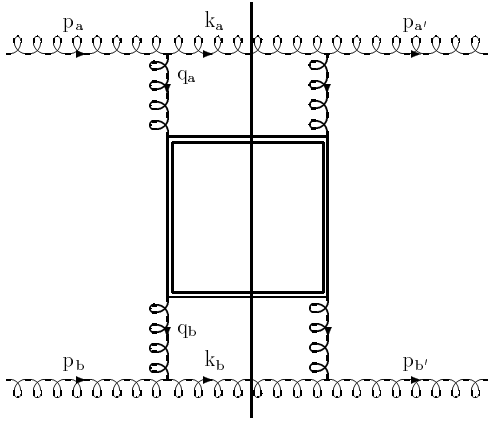}
\vspace*{-9.5cm}
\caption{Two-gluon production with exchange of a BFKL ladder,
represented schematically by the double-lined square.}
\label{fig:six}
\end{figure}

Now, let us go back to the full solution (\ref{solb}). Taking
the inverse Laplace transform (\ref{invd}) of eq.(\ref{sindis}) and using
eq.(\ref{solb}), the discontinuity of the amplitude for singlet exchange is
\beq
{\rm Disc} A^{sing}(\hs,\th=0)\, =\, 16i\,\hs\,\a_s^2\, N_c^2\,\int d^2
q_{a\perp}\, d^2q_{b\perp}\, {1\over q_{a\perp}^2 q_{b\perp}^2} f(q_a,q_b,y)\,
,\label{sdis}
\eeq
with $y = \ln{(\hs/\kt^2)}$ and,
\beq
f(q_a,q_b,y)\, =\, {1\over (2\pi)^2}\, {1\over (\qta^2 \qtb^2)^{1/2}}\,
\sum_{n=-\infty}^{\infty} e^{in\bar\phi}\, \int_{-\infty}^{\infty} d\nu\,
e^{\omega(\nu,n) y}\, e^{i\nu\ln(\qta^2/\qtb^2)}\, ,\label{solc}
\eeq
with $\bar\phi$ the azimuthal angle between the vectors $\qta$ and $\qtb$,
corresponding to the first and the last gluon momenta exchanged on the
ladder (Fig.\ref{fig:six}). $f(q_a,q_b,y)$ may be also written as the inverse
Laplace transform of the solution (\ref{solb}) with respect to $\omega = l-1$.

Next, we consider unitarity and the optical theorem to relate the
total cross section for gluon-gluon scattering with exchange of a one-gluon
ladder (Fig.\ref{fig:four}b) to the forward amplitude for
gluon-gluon elastic scattering with exchange of a
two-gluon ladder (Fig.\ref{fig:five}),
\beq
\hsig_{tot}\, =\, {1\over 2\hs}\, \int d{\cal P} M\, M^{\dagger}\, =\, -{1
\over 2\hs}\, {\rm Disc}\left[i{\overline M}(\hs,\th=0)\right]\,
,\label{optt}
\eeq
with ${\cal P}$ the phase space of eq.(\ref{phsp}), and
where we have used eq.(\ref{disc}). ${\overline M}(\hs,\th=0)$ is the forward
amplitude summed (averaged) over final (initial) colors and helicities.
{}From eq.(\ref{elasgen}) it can be written as
\beq
{\overline M}(\hs,\th)\, =\, {1\over N_c^2-1}\, A^{sing}(\hs,\th)\,
.\label{opts}
\eeq
Using then eq.(\ref{sdis}) and (\ref{opts}), the total cross section for
gluon-gluon scattering (\ref{optt}) in multiregge kinematics is given by
\cite{bal}
\beq
\hsig_{tot}\, =\, {8N_c^2\over N_c^2-1}\, \a_s^2\,\int d^2\kta\,
d^2\ktb\, {1\over \kta^2\ktb^2} f(k_a,k_b,y)\, ,\label{sisig}
\eeq
with $\kta=-\qta$ and $\ktb=\qtb$, as in Fig.\ref{fig:six}. $f(k_a,k_b,y)$
is given by eq.(\ref{solc}) with $\phi = \bar\phi + \pi$ the azimuthal
angle between the vectors $\kta$ and $\ktb$, and $y = \ln{(\hs/\kta\ktb)}$
the rapidity interval between gluons $k_a$ and $k_b$.

As noted after eq.(\ref{om}), the solution (\ref{solb}) or
(\ref{solc}) of the BFKL equation for singlet exchange at $\th = 0$
has no infrared and ultraviolet divergences. Namely, the doubly
logarithmic enhancements of the virtual and real radiative corrections cancel
out, leaving as a residue single logarithms. Accordingly, the expansion of
eq.(\ref{solc}) is finite order by order in $\a_s$. This is due to the
optical theorem (\ref{optt}) which relates the forward scattering amplitude
to the total cross section, for which all the infrared
and ultraviolet logarithmic divergences in the radiative corrections to
the Born amplitude cancel out, order by order in $\a_s$. Besides, as discussed
at the end of sect.~3.3 the coupling constant $\a_s$
in eq.(\ref{bfklb}) must be regarded as fixed, thus the total
cross section (\ref{sisig}) does not depend on a renormalization scale
$\mu$\footnote{Attempts to introduce by hand a running coupling constant in
eq.(\ref{bfklb}) have been made \cite{akms}, but a running $\a_s$
makes eq.(\ref{bfklb}), and so the total cross section
(\ref{sisig}), infrared sensitive.}. Finally, we note that because of the
leading singularity (\ref{asyso}) at $l=1+A$,
the growth of the total cross section, $\hsig_{tot}\, =\,\hs^A$,
violates the Froissart unitarity bound $\s_{tot} \propto
\ln^2{s}$ \cite{pdb}.

\subsection{The total parton cross section}
\label{sec:tresix}

Summarizing the conclusions of sect.\ref{sec:trefive},
the parton cross section for the production of two gluons, resummed to all
orders of $\a_s\ln{(\hs/\kt^2)}$, is obtained from
eq.(\ref{solc}) and (\ref{sisig}),
\begin{equation}
{d\hat\sigma_{gg}\over d^2\kta d^2\ktb}\ =\
\biggl[{C_A\alpha_s\over \kta^2}\biggr] \,
f(k_a,k_b,y) \,
\biggl[{C_A\alpha_s\over \ktb^2}\biggr] \ ,
\label{cross}
\end{equation}
with $y = y_a-y_b$ and $f(k_a,k_b,y)$ the inverse Laplace transform
of the singlet solution $f_l(k_a,k_b)$ (\ref{solb}),
\beq
f(k_a,k_b,y)\, =\, \int {d\omega\over 2\pi i}\, e^{\omega y}\,
f_l(k_a,k_b)\, ,\label{invl}
\eeq
with $\omega = l-1$.
Eq.(\ref{cross}) is written as a convolution of the singlet solution
(\ref{invl}) with the parton-production vertices on each side of the
rapidity interval. The structure
of eq.(\ref{cross}) is generic and valid for any scattering process in
multiregge kinematics (cf. Appendix C).

First, we consider the limit $\a_s y\ra 0$ in eq.(\ref{cross}), for which all
the real and virtual
radiative corrections vanish, and the singlet solution (\ref{invl})
reduces to the inhomogeneous term of the BFKL equation,
\beq
f(k_a,k_b,y)|_{{\cal O}(\a_s^0)}\, =\, \delta(\kta^2-\ktb^2)\,\delta(\phi-\pi)
\, ,\label{corr}
\eeq
i.e. at the Born level only two partons are produced, and they are
balanced in $\kt$ and back-to-back in $\phi$, and
eq.(\ref{cross}) reduces to the Born parton cross section (\ref{yelas}).

By integrating eq.(\ref{cross}) over the azimuthal angle only the $n= 0$
contribution to the eigenvalue (\ref{om}) survives
\beq
{d\hat\sigma_{gg}\over d\kta^2 d\ktb^2}\ =\ {C_A^2\alpha_s^2\over
4\kta^3\ktb^3}\, \int_{-\infty}^{\infty} d\nu\, e^{\omega(\nu,n=0) y}\
e^{i\nu\ln(\kta^2/\ktb^2)}\, .\label{crossb}
\eeq
In the asymptotic limit of large rapidities, we may
approximate the eigenvalue (\ref{om}) with its expansion (\ref{omexp}) at
small $\nu$, and perform the integral over $\nu$ by the saddle-point method,
\beq
{d\hat\sigma_{gg}\over d\kta^2 d\ktb^2}\ =\ {\pi C_A^2\alpha_s^2\over
4\kta^3\ktb^3}\, {e^{Ay}\over \sqrt{\pi B y}}\, \exp\left(-{\ln^2(\kta^2/
\ktb^2)\over 4By}\right)\, ,\label{crs}
\eeq
with $A$ and $B$ given in eq.(\ref{scoef}). Eq.(\ref{crs}) shows a
diffusion pattern, namely it has the form of a Gaussian
distribution in $\ln(\kta^2/\ktb^2)$ with the peak positioned where the partons
are balanced in $\kt$ and with a width growing with $y$. This is not
accidental, since the BFKL equation may be written as a diffusion equation,
with diffusion rate $\ln(\kta^2/\ktb^2)\sim y^{1/2}$ (cf. Appendix D).

Let us go back to eq.(\ref{crossb}), and perform the integration over the
parton transverse momenta above a cutoff $\ptm$.
As noted at the end of sect.~3.3 and \ref{sec:trefive}, in the BFKL
formalism the coupling constant is fixed. Accordingly, for all of the partons
produced in the gluon ladder the coupling must be evaluated at a single scale.
Typical scales are the transverse momenta $\kta$ and $\ktb$
of the partons with respect to which the ladder is being resummed, and their
cutoff $\ptm$. Choosing as a scale any combination of them is theoretically
equally valid, and anyway we expect from eq.(\ref{crs}) that different choices
do not make a large difference within the BFKL approximation. Here
we choose $\a_s=\a_s(\ptm^2)$ \cite{MN}, since it allows us to perform
analitically the integrations over the transverse momenta, thus obtaining
the total parton cross section,
\beq
\hat\sigma_{gg}\, =\, {C_A^2\alpha_s^2\over 4\ptm^2}\,
\int_{-\infty}^{\infty} d\nu\, {e^{\omega(\nu,n=0) y}\over \nu^2+{1\over 4}}\,
.\label{mnt}
\eeq
Approximating then the eigenvalue (\ref{om}) with its expansion (\ref{omexp})
at small $\nu$, and performing the integral over $\nu$, we obtain the
large-$y$ asymptotics of the total cross section,
\beq
\hat\sigma_{gg}\, =\, {\pi C_A^2\alpha_s^2\over 2\ptm^2}\, {e^{4\ln2 N_c
\a_s y/\pi}\over \sqrt{7\zeta(3) N_c \a_s y/2}}\, .\label{hurr}
\eeq
Eq.(\ref{crs}) and (\ref{hurr}) show the
exponential growth in $y$, corresponding to a power-like growth in $\hs$,
characteristic of the pomeron solution (\ref{asyso}).
Dividing eq.(\ref{mnt}) or (\ref{hurr}) by eq.(\ref{yelas}), with $\th$
integrated over the transverse-momentum cutoff $\ptm^2$, we obtain the
$K$-factor, i.e. the growth rate of the total cross section due to the
radiative corrections,
\beq
K\, =\, {1\over 2\pi}\, \int_{-\infty}^{\infty} d\nu\, {e^{\omega(\nu,n=0) y}
\over \nu^2+{1\over 4}}\, \simeq\,
{e^{4\ln2 N_c \a_s y/\pi}\over \sqrt{7\zeta(3) N_c \a_s y/2}}\,
,\label{k}
\eeq
\begin{figure}[hbt]
\vspace*{-9.5cm}
\hspace*{0.5cm}
\epsfxsize=15cm \epsfbox{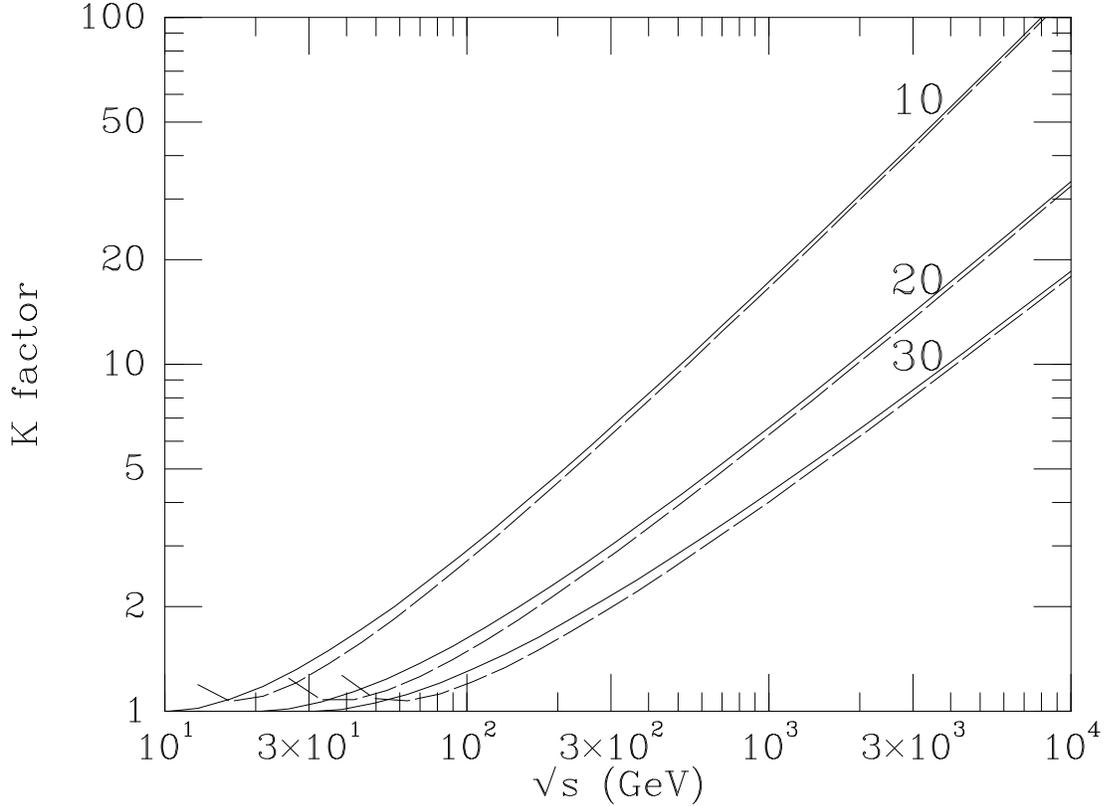}
\vspace*{-0.5cm}
\caption[]{
$K$-factor as a function of
$\sqrt{\hs}$ at different values (in GeV) of
$\ptm$. The solid curves are obtained by doing numerically the integral over
$\nu$ in eq.(\ref{k}), while the dashed curves correspond to the asymptotic
expression on the right-hand side of eq.(\ref{k}).
}
\label{fig:kfact}
\end{figure}
with $y =\ln(\hs/\ptm^2)$. In fig.\ref{fig:kfact} we plot the $K$-factor as a
function of the parton center-of-mass energy $\sqrt{\hs}$ at different values
of the transverse-momentum cutoff $\ptm$. We scale $\a_s=\a_s(\ptm^2)$ from
$\a_s(m_Z^2)=0.12$ using the one-loop evolution with five flavors.
Even though the power of $\hs$
in the $K$-factor is not very large (at $\ptm = 20$
GeV, the power is $\sim 0.4$), the $K$-factor grows quickly
\footnote{We note that the dashed curves approach
the solid ones on the plot logarithmic scale, since the relative error tends
to zero at large $\hs$, however on a linear-scale plot the distance between
the curves increases, since the absolute error increases at large $\hs$
\cite{DDS}.}.

\section{Two-jet production at large rapidities}
\label{sec:four}

In this section inclusive jet production is considered as an
experimental signature of the BFKL theory discussed in sect.~3.
The original proposal of Mueller and Navelet \cite{MN} of linking the growth
rate of the inclusive two-jet production to the $K$-factor of the total
parton cross section
is recalled, and its modifications to fixed-energy colliders are discussed;
finally, the jet-jet decorrelation in transverse momentum is proposed as a
probe of the BFKL dynamics, and an assessment is made of the phenomenological
importance of higher-order and
next-to-leading logarithmic corrections to the BFKL formalism.

\subsection{The Mueller-Navelet jets}
\label{sec:fourone}

In order to analyze two-jet production in a way that resembles
the configuration assumed in the BFKL theory, we select all the jets in
the event above a transverse momentum cutoff $\ptm$, and rank them by their
rapidity, i.e. we tag the two jets with the largest and smallest rapidity and
observe the distributions as a function of these two tagging jets (in the
standard hadronic-jet analysis jets are
ranked by their transverse energy \cite{EKS}-\cite{ggk}). Since
the jet production is inclusive, the distributions are affected by
the hadronic activity in the rapidity interval $y$ between the tagging
jets, whether or not these hadrons pass the jet-selection criteria.

In order to obtain the inclusive two-jet production we must convolute
eq.(\ref{cross}) with parton densities, whose dependence on the parton
momentum fractions is given by eq.(\ref{ykin}), with $k_0=k_a$ and
$k_{n+1}=k_b$.
Our goal is to examine the parton process (\ref{cross}) at growing values
of $y$, to check if indeed it shows the exponential growth in $y$ of the
pomeron solution. At the same time it would be convenient to keep the $x$'s
fixed, in order to
disentangle the eventual dynamical rise due to eq.(\ref{cross}) from kinematic
variations induced by the parton densities. Thus, Mueller and Navelet \cite{MN}
have proposed to measure two-jet production at fixed $x$'s and different
values of
$y$. This implies that the hadron center-of-mass energy $\sqrt{s}$ has to be
risen proportionally with the parton center-of-mass energy $\sqrt{\hs}$, i.e.
with the exponential of the rapidity interval $y$ between the tagging jets.

We need a factorization formula which relates the parton cross section
(\ref{cross}) to jet production. The
factorization formula for the two-jet production cross section in terms of the
jet rapidities and transverse momenta has been given in general in
eq.(\ref{jetfac}). The parton cross section for the production of $n+2$
partons is
\beq
d\hat\sigma_{ij}\, =\, {1\over 2\hs}\, d{\cal P}_{n+2}\, |M_{ij}|^2\, ,
\label{npart}
\eeq
with the differential phase space (\ref{phsp}) for the production of $n+2$
partons. Using the light-cone momentum conservation
in the exact kinematics (\ref{nkin}), and eq.(\ref{npart}), the two-jet
production cross section (\ref{jetfac}) becomes
\bea
& & {d\sigma\over d^2\kta d^2\ktb dy_a dy_b}\, =\label{nexf}\\
& & \sum_{n=0}^{\infty}\,\int
\left(\prod_{l=1}^n \int {dy_l\, d^2 k_{l\perp}\over 4\pi
(2\pi)^2}\right)\, \sum_{ij}\, x_Af_{i/A}(x_A,\mu^2)\, x_Bf_{j/B}(x_B,\mu^2)\,
{|M_{ij}|^2\over 16\pi^2\hs^2}\, \delta^2\left(\sum_{l=0}^{n+1} k_{l\perp}
\right)\, ,\nn
\eea
where the $n=0$ term reproduces eq.(\ref{jetlo}).
In the large-$y$ limit the phase space for the production of $n+2$ partons
is given by eq.(\ref{mrnps}), where the approximate kinematics (\ref{ykin})
is used, and the parton momentum fractions fix the jet rapidities (cf.
eq.(\ref{jac})). Using
eq.(\ref{mrnps}) and (\ref{npart}) into eq.(\ref{nexf}) we obtain the
two-jet production cross section in multiregge kinematics,
\beq
{d\sigma\over d^2\kta d^2\ktb dy_a dy_b}\, =\,
x_A^0 f_{eff}(x_A^0,\mu^2)\, x_B^0 f_{eff}(x_B^0,\mu^2)\,
{d\hat\sigma_{gg}\over d^2\kta d^2\ktb}\, ,\label{mrfac}
\eeq
with the parton cross section as in eq.(\ref{cross}). In eq.(\ref{mrfac})
we have used again the effective parton density (\ref{effec}) since the
leading contribution to two-jet production in multiregge kinematics always
comes from the exchange of a gluon ladder in the $\th$ channel, the only
difference between subprocesses with initial-state quarks or gluons being
the different color strength
in the jet-production vertices. From eq.(\ref{ykin}) and (\ref{mrfac}) we can
now easily derive the factorization formula at fixed parton momentum fractions,
\beq
{d\sigma\over dx_A^0 dx_B^0 d^2\kta d^2\ktb}\, =\,
f_{eff}(x_A^0,\mu^2)\, f_{eff}(x_B^0,\mu^2)\,
{d\hat\sigma_{gg}\over d^2\kta d^2\ktb}\, .\label{mnfac}
\eeq
The value of the factorization scale $\mu$ in the parton densities is
arbitrary, and since from the collinear-factorization standpoint
(cf. the Introduction) eq.(\ref{cross}) has the same accuracy as a LO
calculation, the dependence of the jet-production rate (\ref{mnfac}) on $\mu$
is maximal. In eq.(\ref{mnfac}) Mueller and Navelet fix  the factorization
scale at the transverse momentum cutoff, $\mu = \ptm$, in order to take the
parton densities out of the integration over the jet transverse momenta.
Then the jet-production rate at fixed $x$'s may be directly related to the
total parton cross section (\ref{mnt}),
\beq
{d\sigma\over dx_A^0 dx_B^0}\, =\, f_{eff}(x_A^0,\ptm^2)\,
f_{eff}(x_B^0,\ptm^2)\, \hat\sigma_{gg}\, .\label{mnx}
\eeq
The total parton cross section at the Born level, in the large-$y$ limit,
is given by eq.(\ref{yelas}), with $\th$
integrated over the transverse-momentum cutoff $\ptm^2$. We use it, with
the parton momentum fractions (\ref{largeyx}), in the factorization formula
(\ref{mnx}) to compute
the two-jet production rate at fixed $x$'s, at the Born level.
Normalizing then the two-jet production rate (\ref{mnx}), with
$\hat\sigma_{gg}$ given by eq.(\ref{mnt}),
to the one at the Born level the parton densities cancel
out. Thus, the $K$-factor of two-jet production at fixed $x$'s, i.e. the
growth rate due to the radiative corrections, coincides with the $K$-factor
of the total parton cross section (\ref{k}) \cite{MN}. Then Fig.\ref{fig:kfact}
implies that, could we perform such a measurement and were the BFKL
approximation correct, we should see a large enhancement in the data
with respect to the Born-level estimate of two-jet production at fixed $x$'s.

\subsection{Two-jet production at fixed-energy colliders}
\label{sec:fourtwo}

In sect.\ref{sec:fourone} we have discussed how, in a hadron collider at
variable center-of-mass energy,
the growth rate of two-jet production at fixed parton momentum fractions
may be related to the growth rate of the total parton cross section due to
the BFKL pomeron. However, this measurement
may be difficult to implement experimentally,
since we do not dispose nowadays of a variable-energy collider where such a
{\it ramping run} experiment may be performed\footnote{A {\it ramping run}
experiment could be performed in DIS at the HERA electron-proton collider
\cite{muel}, by varying the electron-energy loss, but it is not clear whether
the rapidity span in jet production in DIS at HERA is large enough to show
the enhancement effect in the $K$-factor.}.

In this section we consider
then two-jet production at fixed $\sqrt{s}$ as a function of the jet
rapidities (\ref{mrfac}). Since at fixed $\sqrt{s}$ and rapidities, the
$x$'s grow linearly with the jet transverse momenta (\ref{ykin}), the
integration over $\ptp$ in the jet production rate will
entail a varying contribution from the parton densities, which we cannot
avoid. Our goal, however, is to examine the parton dynamics and not the
parton densities, so we may at least fix, or integrate out, the rapidity boost
$\bar y=(y_a+y_b)/2$ (cf. sect.\ref{sec:twotre}), since
the parton dynamics (\ref{inv}) and (\ref{invb}) does not depend on it.
Then we rewrite the factorization formula (\ref{mrfac}) as
\beq
{d\sigma\over dy d{\bar y} d\kta^2 d\ktb^2 d\phi}\, =\,
x_A^0 f_{eff}(x_A^0,\mu^2)\, x_B^0 f_{eff}(x_B^0,\mu^2)\,
{d\hat\sigma_{gg}\over d\kta^2 d\ktb^2 d\phi}\, ,\label{dsfac}
\eeq
where the parton cross section is, from eq.(\ref{cross}) and (\ref{invl}),
\begin{equation}
{d\hat\sigma_{gg}\over d\kta^2 d\ktb^2 d\phi}\, =\, {9\alpha_s^2
\over 4\pi\kta^3 \,\ktb^3}\, \sum_{n=-\infty}^{\infty} e^{in(\phi-\pi)}\,
\int_0^{\infty} d\nu e^{\omega(n,\nu)\, y}
\cos\left(\nu \, \ln{\kta^2 \over \ktb^2} \right)\, .\label{mini}
\end{equation}
First, we consider the evaluation of the $K$-factor at fixed $\sqrt{s}$, as
a function of $y$. In sect.\ref{sec:twofive} (cf. Fig.\ref{fig:two})
the two-jet production cross section
$d\s/dy\,d\bar y$ as a function of $y$, at $\bar y=0$, has been computed at
the Born level at the energies of the Tevatron and LHC colliders.
In ref.\cite{DS} (cf. also ref.\cite{DDS} and \cite{stir})
the resummation of the radiative corrections, using eq.(\ref{dsfac}) and
(\ref{mini}), has been considered. At Tevatron energies the corrections
turn out not to be appreciably different from the Born calculation, while at
LHC energies they are substantially larger than the Born estimate, but anyway
much smaller than what expected from the calculation of the $K$-factor at
fixed $x$'s. This is due to the
parton-density contribution to eq.(\ref{dsfac}).
As we know from the analysis of sect.\ref{sec:twofive} and Fig.\ref{fig:two},
the parton densities fall monotonically as $y$ increases. However, the decrease
in phase space is more pronounced for the resummed two-jet cross section than
for the one at the Born level. In order to see it, let us consider the
behavior of the factorization formulae (\ref{dsfac}) and (\ref{logen}) as
$x \ra 1$. The parton densities have a
$(1-x)^{\beta}$ behavior, with $\beta$ positive, as $x \ra 1$.
The $x$'s grow linearly with the jet transverse momenta,
according to eq.(\ref{largeyx}) and (\ref{ykin}). Trasforming
then the integration over the transverse momenta into an integration over the
$x$'s, we see that the resummed two-jet production (\ref{dsfac}) goes like
$(1-x)^{2(\beta+1)}$, as $x\ra 1$, while the one at the Born level
(\ref{logen}) goes like $(1-x)^{2\beta+1}$. Accordingly, the corresponding
$K$-factor falls like $(1-x)$ as $y$ increases \cite{DDS}-\cite{stir}.

\begin{figure}[hbt]
\vspace*{-5.cm}
\hspace*{-0.5cm}
\epsfxsize=15cm \epsfbox{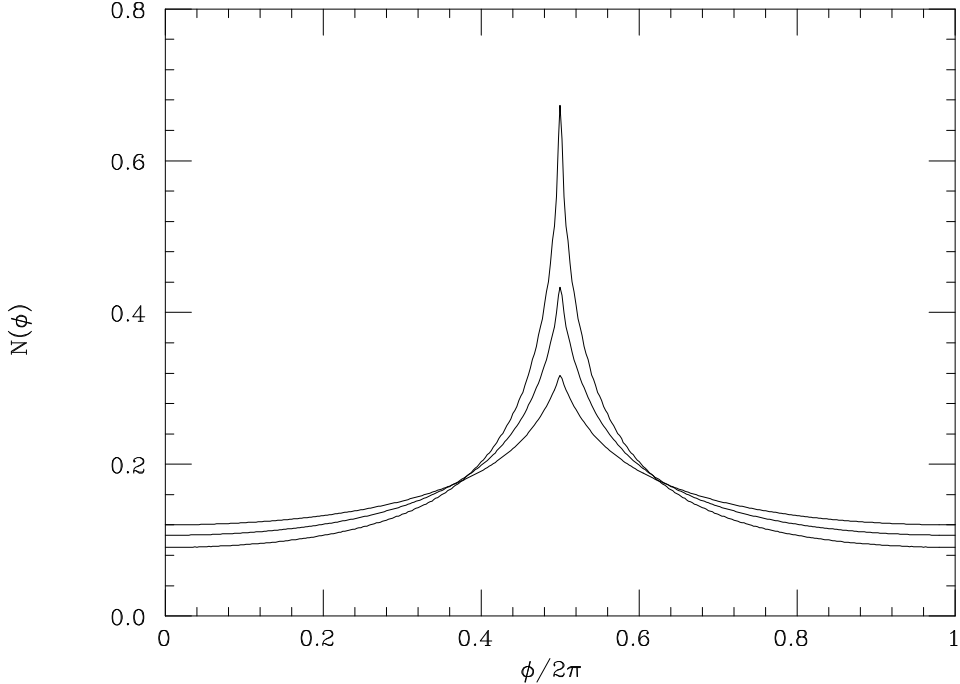}
\vspace*{-6.cm}
\caption[]{$\phi$ distribution at $\sqrt{s}=1.8$ TeV, normalized to the
uncorrelated cross section
$d\sigma/dy d{\bar y}$. From top to bottom, relative to the peak, the solid
lines are the $\phi$ distributions at $y$= 5, 6 and 7.}
\label{fig:phid}
\end{figure}

Another feature of the parton cross section in the BFKL formalism (\ref{cross})
is the jet-jet decorrelation in transverse momentum $\ptp$ and azimuthal angle
$\phi$. Eq.(\ref{cross}) has embodied the correct transverse-momentum
conservation, and at the Born level (\ref{corr}) only two partons, balanced in
$\ptp$ and back-to-back in $\phi$, are produced. However,
in the large-$y$ limit the leading contribution to the eigenvalue (\ref{om}) in
eq.(\ref{cross}) comes from the $n=0$ term (\ref{omexp}), for which the
partons are uncorrelated in $\phi$; and also the correlation
in $\ptp$ fades away as $y$ increases (\ref{crs}). So
eq.(\ref{mini}) should allow us to go from one end (total correlation at
the Born level) to the other (total decorrelation at asymptotically large $y$)
and examine how the
decorrelation in $\ptp$ and $\phi$ increases with $y$. This analysis has
been performed for the $\ptp$-decorrelation in ref.\cite{DDS}, \cite{DS}, and
for the $\phi$-decorrelation in ref.\cite{DS}, \cite{stir}, \cite{DDS3}.
{}From ref.\cite{DS} we report in Fig.\ref{fig:phid} the $\phi$ distribution
at the Tevatron energy as a function of $y$,
at $\bar y=0$. The jet transverse momenta are integrated above $\ptm=20$ GeV,
and the distribution $N(\phi)$ is obtained from the two-jet production rate
$d\sigma/dy d{\bar y} d\phi$ (\ref{dsfac}), normalized to the uncorrelated
one $d\sigma/dy d{\bar y}$. The renormalization and
factorization scales are set to $\mu^2=\kta\ktb$.

The plot of Fig.\ref{fig:phid} may be understood in terms of the BFKL-ladder
picture. As the rapidity interval $y$ between the tagging jets is
increased, along the gluon ladder more and more partons are produced which
decorrelate the tagging jets in $\ptp$ and $\phi$.
The decorrelation in $\ptp$ and $\phi$ as $y$ is increased has been indeed
observed in two-jet production \cite{dzero}, \cite{kim}.
However, it is still to be established whether the interpretation of this
phenomenon in terms of the BFKL ladder is correct, since
the decorrelation may be also produced via the collinear emission
in the parton-density evolution \cite{kim}.

\subsection{The higher-order corrections}
\label{sec:fourtre}

The jet-jet decorrelation examined in sect.\ref{sec:fourtwo} raises a few
questions. How important are in the jet-jet decorrelation configurations
where three or more jets are produced? On kinematic grounds, we would expect
these configurations to be relevant when the tagging jets are not correlated.
How well are reproduced within the BFKL formalism configurations with three
or more jets? To answer these questions we must examine the kinematics in
greater detail.
{}From eq.(\ref{mrnps}) we know that in the multiregge kinematics
transverse-momentum conservation is correctly taken into account, however in
the light-cone momentum conservation (\ref{ykin}) only the leading term is
kept. At the Born level, where both the produced partons are tagged as jets,
we know that this is a good approximation for $y \gsim 3$ (cf. sect.~2.5).
In this section we examine how good the approximation is when
we consider higher-order corrections to the Born-level configuration.
In order to do that, since we do not dispose yet of
complete next-to-leading logarithmic corrections to the BFKL formalism
\cite{lf}, we use the knowledge of the exact three-parton kinematics and
matrix elements \cite{exact}.

\begin{figure}[hbt]
\vspace*{-5.cm}
\hspace*{0.5cm}
\epsfxsize=15cm \epsfbox{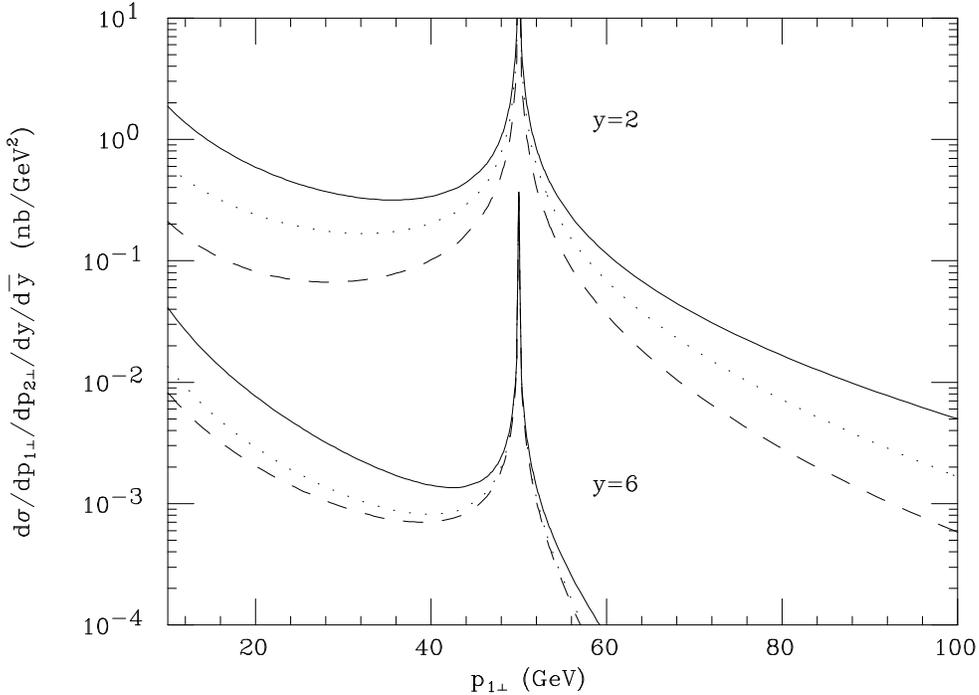}
\vspace*{-6.cm}
\caption{$\ptp$ distribution of the forward jet in rapidity, with the
transverse momentum of the trailing jet fixed at 50 GeV, at ${\bar y} = 0$
and at $y=2$ and 6. The jet-cone size $R_{cut}$ is fixed at 0.7.}
\label{fig:pt}
\end{figure}

In sect.\ref{sec:tretwo} we have noted that the exact matrix elements for
three-gluon production in gluon-gluon scattering (\ref{parke}) reduce in the
limit of strong-rapidity ordering (\ref{mreg}) and (\ref{invb}) to the
ones computed through the BFKL formalism (\ref{trgsq}). To examine the
accuracy of this approximation as well as of the one on the light-cone momentum
conservation (\ref{ykin}), we report from ref.\cite{DDS2}
the contribution of the three-parton configurations to the
$\ptp$ decorrelation. In Fig.\ref{fig:pt} the $\ptp$ decorrelation
is plotted at $\sqrt{s}=1.8$ TeV as a function of the transverse momentum
$p_{1\perp}\equiv\kta$ of
the forward jet in rapidity, at a fixed value of the transverse momentum
$p_{2\perp}\equiv\ktb=50$ GeV of the trailing jet in rapidity. The solid
curves are
computed through the large-$y$ parton cross section (\ref{thpxs}) and
kinematics (\ref{ykin}), using the factorization formula (\ref{dsfac}); the
dotted curves are computed like the solid ones, but using instead the exact
kinematics (\ref{nkin}) for three-parton production; the dashed curves are
computed through the exact matrix elements \cite{exact} and kinematics
(\ref{nkin}), using the three-parton contribution to the exact factorization
formula (\ref{nexf}). All the curves show an infrared divergence at
$\kta\simeq\ktb$, where the third parton becomes soft. In the exact
calculation also collinear divergences are present, when two of the
final-state partons become collinear. These are disposed of by discarding
configurations\footnote{In a complete ${\cal O}(\a_s^3)$ calculation
these configurations would be counted as two-jet events with $\kta=\ktb$,
however they may be neglected in our analysis, since we are only interested
in the modifications induced by the third parton to the large-$y$ kinematics
when $\kta\ne\ktb$.}
where the distance $R=\left[(y_i-y_j)^2+(\phi_i-\phi_j)^2
\right]^{1/2}$ between two partons on the lego plot is smaller than the
jet-cone size $R_{cut}$.

{}From Fig.\ref{fig:pt} we see that the error in using the large-$y$
approximation grows with the imbalance in transverse momentum of the tagging
jets. The dotted curves, which are theoretically inconsistent, are plotted
just to show that while at small $y$'s the error is distributed between the
approximation on the matrix elements and the one on the parton densities,
at large $y$'s most of the error comes in using the
large-$y$ approximation (\ref{ykin}) in the parton densities. To understand it,
we recall that in the three-gluon cross section (\ref{thpxs}) the
intermediate gluon $k_1$ is produced with equal probability over
the rapidity range determined by the tagging jets.
This is also true at the hadron level if we neglect the contribution of the
intermediate gluon to the light-cone momentum conservation (\ref{ykin}).
However, we see from the exact kinematics (\ref{nkin}) that this can be a
bad approximation if $k_{1\perp}$ is not small, particularly when the
intermediate gluon is close in rapidity to the tagging jets. In this case
using the exact kinematics in the parton densities
produces a large suppression, so the rapidity range that the intermediate
gluon may effectively span is reduced substantially.  The
BFKL (solid) curve neglects this effect, and
so it grossly overestimates the cross section\footnote{We have found that even
at LHC energies this discrepancy is not negligible. For example, for the
configuration of Fig.\ref{fig:pt} with $y=10$, $\kta=20$ GeV, $\ktb=50$ GeV,
$R_{cut}=0.7$, the BFKL curve overestimates the exact calculation by almost
a factor 2.}. However, at $\kta\simeq\ktb$
the transverse momentum of the intermediate parton is
small, so its contribution to the $x$'s in (\ref{nkin}) can be safely
neglected.

This entails that the BFKL approximation works at its best when
the tagging jets are balanced in $\ptp$ and back-to-back in $\phi$, and may
have large errors, even at large $y$'s, when the jets are not correlated.
This picture suggests that the tails in the $\phi$ distributions
of Fig.~\ref{fig:phid} may be overestimated, in other words that the $\phi$
decorrelation given by the BFKL approximation is too strong. This seems to
be confirmed by the preliminary data of the D0 Collaboration \cite{kim}.

A remedy to this problem is to use the knowledge of the exact three-parton
configurations and the ambiguity in the definition of
rapidity in the BFKL approximation. We identify the rapidity
$y_{BFKL}=\ln(\hs/\kt^2)$ in the BFKL ladder, with $\kt^2$ a typical
transverse-momentum scale, with the kinematic rapidity $y=y_a-y_b$. However,
the BFKL formalism being a leading logarithmic approximation, $y_{BFKL}$
is defined only up to a factor which is subleading at large rapidities.
In ref.\cite{DDS2} this ambiguity has been exploited in order to define an
effective rapidity $\hat y$ which differs from the kinematic one $y$ by
subleading terms, and which keeps into account the exact three-parton
configurations described above. Using $\hat y$, instead of $y$, in the BFKL
resummation for the $\phi$ decorrelation \cite{DDS3}
has yielded a much better agreement with the data \cite{kim}.
However, this is only a phenomenological
improvement, since from the theoretical point of view to use $y$ or $\hat y$
in the BFKL resummation is equally valid.

\section{Conclusions}

In these lectures we have analysed the BFKL formalism, which resums the
leading logarithmic contributions, in $\ln(\hs/\th)$, to the radiative
corrections to two-parton processes, and we have applied it to inclusive
two-jet production at large rapidity intervals.

After sketching the parton-model and factorization ideas in the Introduction,
all the ingredients necessary to cook up two-jet production at leading
order in $\a_s$ are listed in sect.~2; then the
two-parton dynamics is analysed in detail in the large-$y$ limit
(sect.~2.4) and it is shown that the large-$y$ approximation to
two-jet production agrees well with the exact calculation for $y\gsim 3$.

In sect.~3 the tree-level three-gluon amplitudes (\ref{treg}) and the one-loop
corrections to the two-gluon amplitudes (\ref{vird}) are computed in
the large-$y$ limit, and an ansatz is made on the elastic
(\ref{vtchan}) and the multigluon (\ref{ngluon}) amplitudes in the leading
logarithmic approximation; its self-consistency is shown by using the
multigluon amplitudes to evaluate the elastic amplitude for gluon-gluon
scattering (\ref{vtchan}). In evaluating the elastic amplitude, the BFKL
equation (\ref{bfkle}) which describes the evolution in transverse momentum
of the gluon ladder exchanged in the $\th$ channel
is introduced. The solution of the equation for color-singlet exchange
at $\th=0$ (\ref{solb}), corresponding to the exchange of a (cut) BFKL
pomeron, is then related to the total cross section for gluon-gluon
scattering in the large-$y$ limit (\ref{sisig}). The analytic structure of
the total cross section is examined in sect.~3.8, and its growth rate
with $\hs$, due to the BFKL pomeron, is stressed in Fig.~9.

In sect.~4 the BFKL formalism is applied to estimate two-jet inclusive
production at large $y$'s, starting with Mueller-Navelet proposal which
relates the growth rate of two-jet production at fixed $x$'s (\ref{mnx})
to the one of the total parton cross section; in sect.~4.2 then two-jet
production at fixed-energy colliders at large $y$'s is considered, and the
jet-jet decorrelation in transverse momentum as a function of $y$ is
proposed as a signature of BFKL dynamics. Finally, in sect.~4.3 knowledge
of the exact next-to-leading order kinematics is used to improve the
phenomenological analysis of the jet-jet decorrelation.

Even though the parton dynamics
described by the BFKL formalism is rather sophisticated, a few caveats are
in order in dealing with its applications to jet production: the formalism
has, from the
collinear-factorization standpoint (cf. sect.~1), the same valence as a
leading-order calculation, accordingly it has the maximal dependence on the
renormalization and factorization scales; the rapidity interval to use in the
BFKL resummation is defined only up to subleading terms, which may be quite
important at the energies of the Tevatron and LHC colliders
(cf. sect.\ref{sec:fourtre}); the jet structure, being a subleading feature
of jet production \cite{DDS2}, cannot be resolved by the BFKL formalism.
These considerations and the results of sect.~4.3 seem to suggest that a
complete calculation of next-to-leading logarithmic corrections to the
BFKL formalism should improve substantially the analysis of
jet physics at large rapidities.

\section*{Acknowledgements}

I am very grateful to James Bjorken for triggering and nurturing my interest
in this topics, to Michael Peskin with whom much of sect.~3 was worked out,
to Carl Schmidt for a fruitful collaboration. I also wish to thank
Joachim Bartels, Lev Lipatov, Alfred Mueller, Mark W\"usthoff, Peter Zerwas
for many useful discussions, and Enrico Predazzi for encouraging me to
write this whole thing up.

\appendix
\section{More on the BFKL equation at $\th=0$}

Here we consider the BFKL equation at $\th = 0$ (\ref{bfkla}), without
making the replacement (\ref{iden}) in the virtual radiative-corrections term.
Substituting the solution (\ref{solb}) into the homogeneous equation associated
to eq.(\ref{bfkla}), we obtain the spectrum of the eigenvalues
\beq
\omega(\nu,n)\, =\, {\a_s N_c\over\pi}\, \left[2\, {\rm Re}\int_0^1 dx\,
{x^{{|n|-1\over 2}+i\nu} \over 1-x} - \int_0^1 dx\, {1\over 1-x}\, -\,
{1\over 2}\,\int_0^1 dx\, {1\over x}\right]\, ,\label{appb}
\eeq
with $x$ defined as in eq.(\ref{xvar}).
The eigenvalue has divergences at $x=0 \lra q^2_{2\perp}=0$, and $x=1 \lra
(q_2-q_1)^2_{\perp} = 0$. In order to regulate them, we require that
\bea
q^2_{2\perp} > \mu^2 &\lra& x > \mu^2/q^2_{1\perp}\, ,\label{regul}\\
(q_2-q_1)^2_{\perp} > \mu^2 &\lra& x < 1-\mu/q_{1\perp}\, ,\nn
\eea
with $\mu^2$ an infrared regulator. Then in the last term of eq.(\ref{appb})
we move the singularity from $q^2_{2\perp}=0$ to $(q_2-q_1)^2_{\perp} = 0$,
\beq
{1\over 2}\,\int_{\mu^2/q_2^2}^{1-\mu/q_2} dx\, {1\over x}\, =\,
\int_{\mu^2/q_2^2}^{1-\mu/q_2} dx\, {1\over 1-x}\, ,\label{appc}
\eeq
and we obtain a spectrum which is well behaving in the integration limits,
\beq
\omega(\nu,n)\, =\, 2\,{\a_s N_c\over\pi}\, {\rm Re}\int_0^1 dx\,
{x^{{|n|-1\over 2}+i\nu}-1 \over 1-x}\, ,\label{appd}
\eeq
and agrees with eq.(\ref{om}). The calculation we have outlined here is
equivalent to the one done in sect.~\ref{sec:trefive}, indeed the
regulation of the infrared divergences that we do in eq.(\ref{regul}) and
(\ref{appc}) is done in sect.~\ref{sec:trefive} by eq.(\ref{iden}).

\section{The eigenvalue $\omega(\nu,n)$}

{}From the properties of the $\psi$ function \cite{abr}, the largest values of
the eigenvalue in eq.(\ref{om}) are at $|n|=0$. In particular for $\nu = 0$
we have
\beq
\omega(\nu=0,n)\, =\, -4{\a_s N_c\over\pi}\,\cdot \left\{ \begin{array}{ll}
\displaystyle -\ln{2} + \sum_{k=1}^m {1\over 2k-1} &
\mbox{for $|n|=2m$}\, ,\\ \displaystyle
{1\over 2}\,\sum_{k=1}^m {1\over k} & \mbox{for $|n|=2m+1$}\,
,\end{array} \right. \label{oma}
\eeq
So we fix $|n|=0$ and look at the dependence of $\omega(\nu,n)$ on $\nu$.
Using the expansion \cite{abr}
\beq
\psi(1+y) -\psi(1)\, =\, \sum_{k=2}^{\infty} (-1)^k\, \zeta(k)\, y^{k-1}\, ,
\eeq
valid for small $y$, with $\zeta(k)$ the Riemann $\zeta$-function, and the
doubling formula,
\beq
2\psi(2y)\, =\, 2\ln{2} + \psi(y) + \psi\left(y+{1\over 2}\right)\, ,
\eeq
we obtain
\beq
\omega(\nu,n=0)\, =\, 2{\a_s N_c\over\pi}\, \left[2\ln{2} + \sum_{k=1}^{\infty}
(-1)^k\, (2^{2k+1}-1)\, \zeta(2k+1)\, \nu^{2k}\right]\, .\label{omex}
\eeq

\section{Inserting pomeron amplitudes on Born diagrams}

We show here a simple way of inserting one or more BFKL ladders on cut
diagrams at the Born level. Let us consider the Born cross section for the
production of gluons $k_a$ and $k_b$, in the large-$y$ limit (\ref{yelas}),
rewritten as
\beq
{d\hat\sigma_{gg}\over d^2\kta d^2\ktb}\, =\, {C_A^2\alpha_s^2
\over 2\kta^2 \,\ktb^2}\, \delta^2(\kta+\ktb)\, .\label{appf}
\eeq
Then we insert the (cut) pomeron off-shell amplitude
$f(k_a,k_b,y)$ (\ref{invl}) on the Born diagram, and assume that the
cross section for the production of two gluons, resummed to all orders of
$\a_s y$ has the form,
\beq
{d\hat\sigma_{gg}\over d^2\kta d^2\ktb}\, =\, {C\alpha_s^2
\over\kta^2 \,\ktb^2}\, f(k_a,k_b,y)\, .\label{appg}
\eeq
The coefficient $C$ here is unknown, and is determined by requiring that
at the Born level eq.(\ref{appg}) reproduces eq.(\ref{appf}).
At the Born level there are no radiative corrections, so the singlet
solution (\ref{solb}) is given by the inhomogeneous term of eq.(\ref{bfklb}),
\beq
f_l(k_a,k_b)\, =\, {1\over 2(l-1)}\, \delta^2(\kta+\ktb)\, ,\label{apph}
\eeq
where as usual $\kta=\qta$ and $\ktb=-\qtb$.
Substituting eq.(\ref{invl}) and (\ref{apph}) into eq.(\ref{appg}), and
comparing with the Born cross section (\ref{appf}), we obtain $C=C_A^2$,
and eq.(\ref{appg}) agrees with the resummed parton cross section
(\ref{cross}).

In the same way we may derive the BFKL corrections to the cross
sections for production of two heavy-quark pairs in photon-photon scattering
\cite{bal}, for forward-jet production in DIS \cite{muel}, for
(Higgs-boson~+~one-jet) production in gluon-gluon scattering \cite{DDS},
provided the respective Born cross sections are known in the large-$y$ limit.

\begin{figure}[hbt]
\vspace*{-6.5cm}
\hspace*{-0.5cm}
\epsfxsize=15cm \epsfbox{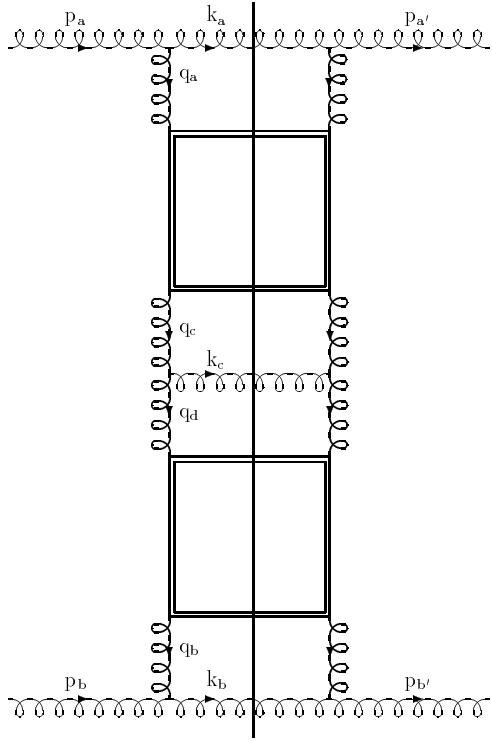}
\vspace*{-5.0cm}
\caption{Three-gluon production with exchange of two BFKL ladders,
represented schematically by the double-lined squares.}
\label{fig:app}
\end{figure}

Now suppose that we want to derive the cross section for the production of
three gluons $k_a$, $k_b$ and $k_c$, with rapidity ordering $y_a \gg y_c \gg
y_b$, assuming that between each pair of gluons $(k_a,k_c)$ and $(k_c,k_b)$ a
BFKL ladder is exchanged \cite{mike}. The corresponding Born cross section
may be derived from eq.(\ref{thpxs}), or from eq.(\ref{appg}) with the singlet
solution (\ref{solb}) corresponding to the emission of a
real gluon in the rapidity interval $y_a-y_b$  (cf. the kernel of
eq.(\ref{bfklb})),
\beq
f_l(k_a,k_b)\, =\, {\a_s C_A\over\pi^2}\, {1\over (k_a-k_b)^2}\,
{1\over 2(l-1)^2}\, .\label{appl}
\eeq
and we obtain
\beq
{d\hat\sigma_{gg\ra ggg}\over d^2\kta d^2\ktb d^2k_{c\perp} dy_c}\, =\,
{C_A^3\alpha_s^3\over 2\pi^2\kta^2\ktb^2 k_{c\perp}^2}\,
\delta^2(\kta+\ktb+k_{c\perp})
\, .\label{appm}
\eeq
where the rapidities of gluons $k_a$ and $k_b$ are as usual fixed by the
parton momentum fractions. Guided by eq.(\ref{appm}), we assume that the
cross section for three-gluon production with the insertion of two BFKL
ladders has the form (Fig.\ref{fig:app}),
\bea
& & {d\hat\sigma_{gg\ra ggg}\over d^2\kta d^2\ktb d^2k_{c\perp} dy_c}\, =
\label{appn}\\ & & D\, {\alpha_s^3\over \kta^2\ktb^2 k_{c\perp}^2}\, \int
d^2q_c d^2q_d\,\delta^2(q_{c\perp}-q_{d\perp}-k_{c\perp})\,
f\left(q_a,q_c,y_{ac}\right)\, f\left(q_d,q_b,y_{cb}\right)\, ,\nn
\eea
with $y_{ac}=y_a-y_c$ and $y_{cb}=y_c-y_b$.
In order to determine the coefficient $D$ we take the singlet solution
at the Born level (\ref{apph}), and insert it into the pomeron amplitudes
$f\left(q_a,q_c,y_{ac}\right)$ and $f\left(q_d,q_b,y_{cb}\right)$.
By comparing then the result with the corresponding Born cross section
(\ref{appm}), we obtain $D = 2C_A^3/\pi^2$. Thus, eq.(\ref{appn}) agrees
with the three-gluon production cross section with exchange of two BFKL
ladders computed in ref. \cite{mike}. This procedure is straightforwardly
generalizable to the production cross section for $n$ rapidity-ordered gluons,
with exchange of a BFKL ladder between each pair of nearest-neighbor gluons.

\section{The diffusion equation}

In order to understand the asymptotic behavior (\ref{crs}) of the parton
cross section, which follows from taking the leading singularity (\ref{asyso})
of the partial-wave solution in the complex plane of $l$, we must consider the
recursive relation (\ref{recur}) at a very
high value of the order parameter $n$ in the iteration procedure.
Accordingly, we take the BFKL equation (\ref{bfklb}) at a high value of $n$,
for which the inhomogeneous term is not important. Then from eq.(\ref{homog})
we obtain
\bea
& & (l-1)\, f_l^{(n)}(q_1,k)\, =\label{diff}\\ & &
{\a_s N_c\over \pi^2} \int d^2q_{2\perp}\, \left[ {f_l^{(n-1)}(q_2,k)
-\,(q_{1\perp}^2/q_{2\perp}^2)\, f_l^{(n-1)}(q_1,k)\over
(q_1-q_2)_{\perp}^2}\, +\, {q_{1\perp}^2 f_l^{(n-1)}(q_1,k)\over q_{2\perp}^2
\left[q_{2\perp}^2 + (q_1 - q_2)_{\perp}^2\right]}\right]\, .\nn
\eea
Next, we take the solution (\ref{solb}), averaged over the azimuthal
angle, and consider only its dependence on $q_{1\perp}$,
\beq
f_l^{(n)}(q_1) \sim (q_{1\perp}^2)^{-1/2} \psi_n(\lambda_1)\, ,\qquad
{\rm with}\quad \lambda_1 = \ln(q_{1\perp}^2/m^2)\, ,\label{difb}
\eeq
and we replace it in eq.(\ref{diff}), in the limit $(n |\ln(q_{1\perp}^2/
q_{2\perp}^2)|) \ll 1$, for which the produced partons are approximately
balanced in $q_{\perp}$. Then we expand the solution (\ref{difb}) taking the
configuration where the partons are balanced in $q_{\perp}$ as a stable point,
\beq
\psi_{n-1}(\lambda_2)\, =\, \psi_{n-1}(\lambda_1)\, +\, {(\lambda_2-
\lambda_1)^2\over 2} {\partial^2 \psi_{n-1}(\lambda_1)\over\partial
\lambda_1^2}\,
,\label{difc}
\eeq
and for the recursive relation (\ref{diff}) we obtain
\bea
(l-1)\, \psi_n(\lambda_1)\, &=&
{\a_s N_c\over \pi} \int_0^{\infty} d\lambda_2 \left[\left( {e^{(\lambda_2-
\lambda_1)/2} - 1 \over |1-e^{\lambda_2-\lambda_1}|}\,+\, {1\over \sqrt{1 +
4e^{2(\lambda_2- \lambda_1)}}}\right)\, \psi_{n-1}(\lambda_1)\right.
\nn\\ &+& \left. {(\lambda_2-
\lambda_1)^2\over 2}\, {e^{(\lambda_2-\lambda_1)/2}\over |1-e^{\lambda_2-
\lambda_1}|}\, {\partial^2 \psi_{n-1}(\lambda_1)\over\partial\lambda_1^2}
\right]\, .\label{difd}
\eea
The integration over the first two terms on the right-hand side yields the
eigenvalue (\ref{om}), of which we take the asymptotic solution (\ref{omexp}).
Neglecting then subleading terms, the recursion relation becomes
\beq
(l-1)\, \psi_n(\lambda_1)\, =\, A\, \psi_{n-1}(\lambda_1)\, +\,
{\a_s N_c\over \pi}\, \int_0^{\infty} dz\, {z^2 e^{z/2}\over e^z -1}\,
{\partial^2 \psi_{n-1}(\lambda_1)\over\partial\lambda_1^2}\, ,\label{dife}
\eeq
with $A$ given in eq.(\ref{scoef}). Using the integral representation
of the Riemannn $\zeta$-function \cite{abr},
\beq
\zeta(x)\, =\, {2^x\over (2^x-1) \Gamma(x)}\, \int_0^{\infty} dt\,
{e^t t^{x-1}\over e^{2t}-1}\, ,\label{difg}
\eeq
and B as given in eq.(\ref{scoef}), and going to continuous values for $n$,
we obtain a diffusion equation,
\beq
(l-1)\, {\partial\psi(n,\lambda) \over\partial n}\, =\, \left[A-(l-1)\right]\,
\psi(n,\lambda)\, +\, B\, {\partial^2 \psi(n,\lambda)\over\partial
\lambda^2}\, .\label{difh}
\eeq
This is a Schr\"odinger-type equation, with $n$ playing the role of time, and
with a constant {\sl potential}, $A-(l-1)$. Therefore the solution is
separable \cite{sch}, $\psi(n,\lambda)=f(n)g(\lambda)$. At
the singularity $l = 1+A$ the diffusion equation is,
\beq
A\, {\partial\psi(n,\lambda) \over\partial n}\, =\,
B\, {\partial^2 \psi(n,\lambda)\over\partial\lambda^2}\, ,\label{difl}
\eeq
with diffusion rate $\lambda\sim n^{1/2}$. Supposing that at
the {\sl time} $n=0$ the wave function has a Gaussian shape,
\beq
\psi(n=0,\lambda)\, =\, {1\over (\sigma^2\pi)^{1/4}}\, \exp\left(-{\lambda^2
\over 2\sigma^2}\right)\, ,\label{difi}
\eeq
we obtain the solution as a wave packet spreading in time \cite{lipat},
in qualitative agreement with eq.(\ref{crs}) if we make the correspondence
$A y\sim n$,
\beq
\psi(n,\lambda)\, \sim\, \left({A\over 4Bn}\right)^{1/2}\,
\exp\left(-{A\lambda^2\over 4Bn}\right)\, ,\label{difj}
\eeq
where we have neglected the initial width, $A\sigma^2 \ll 2Bn$.

\end{document}